\documentclass[longauth]{aa} % for the long lists of affiliations 
\usepackage{graphicx}
\usepackage{txfonts}
\usepackage{color,soul}
\usepackage[dvipsnames]{xcolor}
\newcommand{\msun}{${\rm M}_\odot$}
\newcommand{\lsun}{${\rm L}_\odot$}
\newcommand{\rsun}{${\rm R}_\odot$}

\def\rstar{${\rm R}_\star$}
\def\mstar{${\rm M}_\star$}
\def\lstar{${\rm L}_\star$}

\def\prot{${\rm P}_{rot}$}

\def\rsub{$R_{sub}$}

\def\rdust{$R_{dust}$}
\def\rbrg{$R_{Br\gamma}$}
\def\rco{$R_{CO}$}

\def\macc{$\dot{M}_{\rm acc}$}
\def\lacc{$L_{\rm acc}$}

\def\vsini{$v\sin i$}
\def\prot{${\rm P}_{rot}$}
\def\vrad{${\rm V}_{r}$}

\def\msunyr{${\rm M}_\odot {\rm yr}^{-1}$}
\def\deg{$\degr$}

\def\kms{km~s$^{-1}$}

\def\hei{${\rm HeI}$}

\def\brg{${\rm Br}\gamma$}
\def\pab{${\rm Pa}\beta$}

\def\lacc{L$_{acc}$}
\def\lline{L$_{line}$}

\begin{document} 

   \title{The GRAVITY young stellar object survey}
   \subtitle{XV. The star--disk interaction region of the T Tauri star DO~Tau\thanks{ESO VLTI GTO programs with run ID 110.23TT.002 and 112.25T1.001. Partly based on observations obtained at the Canada-France-Hawaii Telescope (CFHT) which is operated by the National Research Council (NRC) of Canada, the Institut National des Sciences de l'Univers of the Centre National de la Recherche Scientifique (CNRS) of France, and the University of Hawaii. The observations at the CFHT were performed with care and respect from the summit of Maunakea.} } 

\author{GRAVITY Collaboration: K. Perraut\inst{1} \and J. Bouvier\inst{1} \and H. Nowacki\inst{1,2} \and A. Sousa\inst{1} \and M. Houllé\inst{1} \and J.F. Donati\inst{3} \and E. Alecian\inst{1} \and S. Alencar\inst{4} \and M. Audard\inst{5} \and J.-P. Berger\inst{1} \and Y.-I. Bouarour\inst{1} \and E. Bordier\inst{6} \and G. Bourdarot\inst{7} \and A. Carmona\inst{1} \and A. Caratti o Garatti\inst{8} \and C. Dougados\inst{1} \and M. Flock\inst{9} \and R. Garcia-Lopez\inst{10} \and K. Grankin\inst{11} \and \'A. K\'osp\'al\inst{9,12,13} \and L. Labadie\inst{6} \and C. Moutou\inst{3} \and J. Sanchez-Bermudez\inst{14} \and H. Shang\inst{15} \and M. Takami\inst{15} \and A. Amorim\inst{16,17} \and W. Brandner\inst{9} \and Y. Cl\'enet\inst{18} \and R. Davies\inst{7} \and R. Dembet\inst{18} \and A. Drescher\inst{1} \and A. Eckart\inst{6} \and F. Eisenhauer\inst{7} \and M. Fabricius\inst{7} \and H. Feuchtgruber\inst{7} \and N.M. F\"orster-Schreiber\inst{7} \and P. Garcia\inst{19,20} \and E. Gendron\inst{18} \and R. Genzel\inst{7} \and S. Gillessen\inst{7} \and T. Henning\inst{9} \and L. Jocou\inst{1} \and S. Joharle\inst{7} \and P. Kervella\inst{18} \and L. Kreidberg\inst{9} \and S. Lacour\inst{18} \and V. Lapeyr\`ere\inst{18} \and J.-B. Le Bouquin\inst{1} \and D. Lutz\inst{7} \and F. Mang\inst{7} \and T. Ott\inst{7} \and T. Paumard\inst{18} \and G. Perrin\inst{18} \and S. Rabien\inst{7} \and D.C. Ribeiro\inst{7} \and M. Sadun Bordoni\inst{7}\and D. Santos\inst{7} \and J. Shangguan\inst{7} \and T. Shimizu\inst{7} \and C. Straubmeier\inst{6} \and E. Sturm\inst{7} \and L. Tacconi\inst{7} \and F. Vincent\inst{18}}

   \institute{Univ. Grenoble Alpes, CNRS, IPAG, 38000 Grenoble, France\\
   \email{karine.perraut@univ-grenoble-alpes.fr}
\and
Laboratoire Lagrange, Université Côte d'Azur, Observatoire de la Côte d'Azur, CNRS, Boulevard de l'Observatoire, CS 34229, 06304, Nice Cedex 4, France
\and
Univ. de Toulouse, CNRS, IRAP, 14 avenue Belin, 31400, Toulouse, France
\and
Departamento de Fisica - ICEx - UFMG, Av. Antônio Carlos 6627, 30270-901, Belo Horizonte, MG, Brazil
\and
Department of Astronomy, University of Geneva, Chemin Pegasi, 51, Versoix CH-1290, Switzerland
\and
I. Physikalisches Institut, Universit\"at zu K\"oln, Z\"ulpicher Strasse 77, 50937, K\"oln, Germany
\and
Max Planck Institute for Extraterrestrial Physics, Giessenbachstrasse, 85741 Garching bei M\"{u}nchen, Germany
\and
INAF - Osservatorio Astronomico di Capodimonte, Salita Moiariello 16, 80131 Napoli, Italy
\and
Max Planck Institute for Astronomy, K\"onigstuhl 17,
69117 Heidelberg, Germany
\and
University College Dublin (UCD), Dublin, Ireland
\and 
Crimean Astrophysical Observatory, 298409, Nauchny, Republic of Crimea
\and
Konkoly Observatory, HUN-REN Research Centre for Astronomy and Earth Sciences, MTA Centre of Excellence, Konkoly-Thege Miklós út 15-17, 1121, Budapest, Hungary
\and
Institute of Physics and Astronomy, ELTE Eötvös Loránd University, Pázmány Péter sétány 1/A, 1117, Budapest, Hungary
\and
Instituto de Astronom\'ia, Universidad Nacional Aut\'onoma de M\'exico, Apdo. Postal 70264, Ciudad de M\'exico 04510, Mexico
\and
Institute of Astronomy and Astrophysics, Academia Sinica, Roosevelt Rd, Taipei 10617, Taiwan
\and
Universidade de Lisboa—Faculdade de Ciências, Campo Grande, 1749-016 Lisboa, Portugal
\and
LIP, Laboratory for Instrumentation and Experimental Particle Physics, Av. Prof. Gama Pinto 2, Lisbon, 1649-003, Portugal
\and
LIRA, Observatoire de Paris, Université PSL, CNRS, Sorbonne Université, Université Paris Cité, 5 place Jules Janssen, 92195, Meudon, France
\and
Universidade do Porto, Faculdade de Engenharia, Rua Dr. Roberto Frias, 4200-465 Porto, Portugal
\and
CENTRA—Centro de Astrofísica e Gravitação, IST, Universidade de Lisboa, 1049-001 Lisboa, Portugal
}
   \date{Received; accepted}
   
  \abstract
   {Protoplanetary disks around young Sun-like stars are the cradles of the vast majority of detected exoplanets. Probing these disks at multiple spatial scales is key to uncovering how planets form. The inner astronomical unit, the star–disk interaction region, is of utmost importance because most detected exoplanets occupy this zone.
   }
   {We aim to spatially and spectrally resolve the inner disk and star-disk interaction region of the M0.3 T Tauri star DO Tau by combining two complementary techniques. 
   }
   % Methods
   {We used high-resolution near-infrared spectra from CFHT/SPIRou to constrain the magnetospheric star-disk interaction process and optical long-baseline interferometry with ESO VLTI/GRAVITY to determine the sizes of the K-band continuum and Br$\gamma$ line emitting regions. From the SPIRou spectra, we measured the veiling in the YJHK bands along with the equivalent widths of the HeI~$\lambda$1083, Pa$\beta$, and Br$\gamma$ emission lines, from which we estimated the mass accretion rate. We were able to monitor the time variability of these quantities thanks to our long-sequence of observations over about 40 days. We fit the GRAVITY visibilities in the continuum and the differential quantities in the line with geometrical models to obtain the orientation and the size of the inner disk as well as the size and the on-sky displacement of the Br$\gamma$ emitting region.
   }
  % results
   {We derived a mass accretion rate of $\sim$~10$^{-8}$ -- 10$^{-7}$~\msunyr, which confirms that this $\sim$0.5~\msun\ star is a strong accretor. The HI and HeI lines exhibit strong variability on a daily timescale, consistent with the burster classification of DO~Tau derived from its K2 light curve. We report a periodic modulation of the intensity of the redshifted high-velocity wings {of the \brg\ line profile}. The modulation occurs at the rotational period of the star (5.128~d), which suggests the existence of corotating magnetospheric funnel flows. We derived an upper limit of 0.35 on the ratio between the magnetospheric truncation radius and the disk corotation radius, indicative of an ordered unstable accretion regime. The size of the \brg{} line emitting region obtained from GRAVITY is quite small ($R_{Br\gamma}$~=~0.011~au~$\sim$~1.3~\rstar), and it is much smaller than the K-band continuum emitting region ($R_{K}$~=~0.09~au~$\sim$~11~\rstar). Such a compact \brg{} emission region suggests that most of the line flux originates from the magnetospheric accretion region and/or from an inner wind close to the magnetosphere-disk interface. The on-sky displacements of the blue and red \brg\ line velocity channels suggest a rotation pattern of the emitting gas, as they appear to be nearly aligned along the position angle of the disk. The inclination we derived for the inner disk ($\sim$~45-55$^\circ$) differs from that of the outer disk inferred from the ALMA continuum ($\sim$~30$^\circ$). This points toward a misalignment or warp of the outer disk that may originate from the suspected past encounter with the neighboring HV~Tau system.
      }
   {Based on combining high-resolution spectroscopy and long baseline interferometry, we find that the T Tauri star DO~Tau appears to be a strong accretor undergoing magnetospheric accretion in an ordered unstable regime, with a \brg{} line emitting region as compact ($\sim$~0.01~au) as the size of its magnetosphere.
   }

   \keywords{Stars: formation -- Stars: pre-main sequence  -- Accretion, accretion disks -- Stars: circumstellar matter -- Instrumentation: interferometers -- Techniques: high angular resolution -- Techniques: interferometric -- Stars: individual: DO Tau}

 \titlerunning{The star--disk interaction region of the T Tauri star DO~Tau}
 \authorrunning{GRAVITY Collaboration}
    
   \maketitle

\begin{table*}[t]
        \caption{{Stellar parameters of DO~Tau adopted in \citet{Long2019}, \citet{Alcala2021}, and this work. }}
        \label{tab:param}
    \begin{center}
    \begin{tabular}{l c c c l }
    \hline
    \hline
    Parameters & Long et al. (2019) & Alcal\'a et al. (2021) & Values adopted in this work &  References \\
    \hline
    $d$ [pc] & 139 & 138.5 $\pm$ 0.7 & 138.5 $\pm$ 0.7 & (a) \\  [1ex]      
    $M_*$ [M$_\odot$] & 0.59$^{+0.15}_{-0.13}$ & 0.5 & 0.54 $\pm$ 0.07 & (b)\\ [1ex]
    $T_{\rm eff}$ [K] & 3806 & 3694 $\pm$ 104 & 3500 $\pm$ 50 & (c)\\  [1ex]
    $\log g$ & -- & -- & 3.6 $\pm$ 0.1 & (c)\\  [1ex]    
    $v \sin i$ [km/s] &-- & 12.0 $\pm$ 2.2 & 13.0 $\pm$ 0.5 & (c) \\  [1ex]
    P$_{\rm rot}$ [d] &-- & -- & 5.128 $\pm$ 0.003 & (c) \\  [1ex]     
    $i$ [$^\circ$] & -- & -- & 49 $\pm$ 5$^{(*)}$ & this work \\  [1ex]     
    $R_*$ [R$_\odot$] & 1.1 & 1.58 & 1.8 $\pm$ 0.2 & this work \\  [1ex]  
    $L_*$ [L$_\odot$] & 0.23 & 0.4 & 0.44 $\pm$ 0.08 & this work\\  [1ex]
    $\dot{M}_{\rm acc}$ [M$_\odot$/yr] & -- & 1.8 10$^{-8}$ & 10$^{-8}$ -- 10$^{-7}$ & this work\\  [1ex]
    \hline
\end{tabular}
\end{center}
\noindent {\bf Notes.} The last column provides the references for the values adopted in this work: (a) \cite{GaiaDR3}; (b) \citet{Braun2021}; (c) \cite{Donati2026}.

\noindent $^{(*)}$ We assume that the inclination of the star equals the inclination of the inner disk determined from GRAVITY measurements.
\end{table*}

\section{Introduction}
\begin{nolinenumbers}

The accretion of matter from the circumstellar disk onto the central star plays a crucial role in the star and planet formation process. It impacts the properties of the central star, its early evolution, and the large amount of energy radiation from accretion shocks that irradiates the circumstellar disk and the protoplanets, thus drastically affecting their physical and chemical evolution. Star and planet formation and disk evolution are intrinsically linked and hence influence each other. In the magnetospheric accretion scenario, the strong stellar magnetic field of young solar-mass T Tauri stars truncates the inner disk out to a few stellar radii, with the accreted matter being funneled onto the star along the magnetic field lines \citep{Bouvier2007,hartmann2016}. As the accreting gas falls onto the stellar surface at high speed, the resulting accretion shock(s) and funnel flows can be traced through emission lines in the visible and/or in the near-infrared ranges. The interaction between the star and the inner disk also likely plays a role in the launching and the shaping of large-scale jets or disk winds. Probing the inner astronomical unit around young stars is key to understanding accretion-ejection phenomena and planet formation.

At the distance of the nearest star forming regions ($\sim$140~pc), zooming in on such star-disk interaction scales requires a spatial resolution of a fraction of a milliarcsecond (mas). Reaching this spatial resolution recently became possible with near-infrared long-baseline interferometry, {notably through spectro-astrometry}. Using the K-band GRAVITY interferometric instrument \citep{GRAVITY} at the combined focus of the Very Large Telescope Interferometer (VLTI), \cite{Soulain2023, Wojtczak2023}, and \cite{nowacki_2024} were able to spatially resolve the innermost disk regions and derive the sizes of the hydrogen I \brg\ line emitting regions around about ten T Tauri stars. These sizes range from 0.03~au to 0.24~au (i.e., from 6.5 to 52~$R_\odot$). For these objects, the authors also derived on-sky photocenter displacements of the \brg\ line with respect to the continuum of a few hundredths of an astronomical unit, and they traced accretion flows at this spatial scale. Taking benefit from the new hints provided by optical long-baseline interferometry about the sizes and the on-sky photocenter displacements of the emitting regions, \cite{Tessore2023} and \cite{Wojtczak2024} developed two families of radiative transfer models to reproduce the interferometric observations: one based on a pure magnetospheric scenario that reproduces well the observations for low accretors (e.g., TW~Hya or DoAr~44) and a second one including a disk wind on top of the pure magnetospheric scenario to emulate the observations of stronger accretors (e.g., RU~Lup and SCrA~N).

High spectral resolution spectroscopy in the optical and in the near-infrared provides proxies to study the accretion-ejection processes \citep[e.g.,][]{Alencar2012,Sousa2023}, and it allows us to investigate the time variability of the emission line profiles formed at least in part in funnel flows, to measure the continuum excess coming from the disk emission in the near-infrared, to estimate the mass accretion rate from the line fluxes, and to determine the epoch when the accretion shock faces the observer. In addition, using spectropolarimeters such as ESPaDOnS \citep{Donati2003} and SPIRou \citep{Donati2020} at the Canada-France-Hawaii-Telescope (CFHT) allows the magnetic field strength and large-scale topology to be derived.
Hence, high-spectral resolution spectroscopy and high-angular-resolution optical interferometry strongly complement each other. When combined, they provide a powerful framework for drawing a coherent picture of the innermost regions of the protoplanetary disks down to the star-disk interaction region, as already clearly demonstrated by pioneering works \citep{Bouvier2020a, Bouvier2020b, GarciaLopez2020,   nowacki_2024}.

This paper presents the first results of an observing campaign on the strongly accreting T Tauri star DO~Tau that combines VLTI/GRAVITY and CFHT/SPIRou observations. In Sect. 2, we describe the properties of the object. Section 3 details the near-infrared spectroscopic observations and results obtained with SPIRou, and in Sect. 4 we describe the GRAVITY K-band interferometric observations and their modeling. We discuss our findings in Sect. 5. The analysis of the spectro-polarimetric SPIRou datasets to monitor the magnetic field topology is presented in a companion paper \citep{Donati2026}.

\section{DO Tau, a strong accretor}

DO~Tau is a low-mass, M0.3 T Tauri star located in the C7-L1527 subgroup of the Taurus cloud \citep{Herczeg2014}, at a mean distance of 141.81 pc and an estimated age of 2.59 $\pm$ 0.75 Myr \citep{Krolikowski2021}, {which is larger than the age of 0.6$_{-0.4}^{+1.1}$~Myr derived by \citet{Luhman2023}}. The large-scale environment of DO Tau is quite complex, with a number of arc-like structures extending over hundreds to thousands of astronomical units, some of which have been interpreted as the remnants of a stellar encounter with the neighboring HV~Tau system about 0.1 Myr ago \citep{Winter2018}. Probably due to this complex circumstellar environment, the extinction toward the star is quite uncertain. Estimates range from A$_V$~=~0.75~mag \citep{Herczeg2014} to 1.7~mag \citep{Alcala2021}, and even 3.04~mag \citep{Fischer2011}. {In these previous studies, the stellar luminosity \lstar\ ranges between 0.23 and 0.4~\lsun, and the stellar radius \rstar\ between 1.1 and 1.58~\rsun \citep{Long2019,Alcala2021}.}
%Using a distance of 139~pc, \cite{Long2019} derived a luminosity \lstar\ = 0.23~\lsun, a radius \rstar\ = 1.1~\rsun, and a mass \mstar\ =~0.59~\msun. Adopting a distance for DO~Tau in line with the Gaia parallax \citep[138.5~$\pm$~0.7~pc;][]{GaiaDR3}, \cite{Alcala2021} derived for the stellar parameters: \vsini\ = 12.0~$\pm$~2.2~\kms, \teff\ = 3694~$\pm$~104~K, \lstar\ =~0.4~\lsun, and \rstar\ = 1.58~\rsun.
Using ALMA data, \citet{Braun2021} estimated for DO~Tau a dynamical mass \mstar\ = 0.54~$\pm$~0.07~\msun\ from the Keplerian rotation of its circumstellar disk. In this work, we adopted the effective temperature, gravity, stellar rotational period, and \vsini\ determined from the SPIRou spectra \citep{Donati2026}, while we assumed the inclination of the star is the same as the inclination of the inner disk we measured here {(see Sect. 4.2)}. From the stellar inclination, rotational period, and \vsini\ we thus derived a stellar radius $R_*$~=~1.8 $\pm$ 0.2~R$_\odot$, in agreement with the evolutionary tracks of \cite{Feiden2016} including magnetic field, and a stellar luminosity $L_*$~=~0.44~$\pm$~0.2~L$_\odot$. {All these stellar parameters are summarized in Table~\ref{tab:param} for comparison.}

\cite{Dodin2013} reported a strong surface magnetic field for DO Tau, with a longitudinal component amounting to a few hundred gauss in the narrow component of the HeI line profile at 587.6~nm. Such a strong field may truncate the inner disk and lead to magnetospheric accretion onto the star, as suggested by the appearance of a highly redshifted absorption component in the \pab\ line profile \citep{Folha2001}. In contrast, the \hei\ 1083~nm line exhibits a pure P-Cygni profile, whose shape suggests the existence of a stellar wind \citep{Kwan2007}. \cite{Sousa2023} derived a barycentric radial velocity \vrad\ = 16.1~$\pm$~0.1~\kms\ from a series of high-resolution near-infrared spectra obtained with SPIRou \citep{Donati2020}, and estimated the mass accretion rate \macc\ = 2.9 $\pm$ 0.5 10$^{-8}$~\msunyr\ from the \pab\ and \brg\ line fluxes assuming $A_v$ = 0.75 mag. This value is similar to earlier estimates computed from optical and near-infrared line fluxes \citep[e.g., 1.6 10$^{-8}$~\msunyr;][]{Gangi2022}, but an order of magnitude lower than the mass accretion rate derived from the UV-optical excess \citep[1.4 10$^{-7}$~\msunyr\, assuming  A$_V$~=~2.27~mag;][]{Gullbring1998}. The mass accretion rate may thus vary significantly on a timescale of years and perhaps much shorter, as suggested by \cite{Cody2022}'s classification of the source as a burster from its K2 light curve, indicative of non-steady accretion bursts occurring at the stellar surface. With its high optical and near-infrared veiling, large UV and near-infrared excesses, wide and strong emission lines \citep{Alcala2021,Sousa2023}, DO Tau certainly belongs to the class of relatively strong accretors among Class II sources of similar mass.

As for any Class II T Tauri star, DO~Tau is surrounded by a gas-rich and dust-rich circumstellar disk. This disk appears to be compact and continuous, with an outer radius of about 80 au measured in scattered light at 1.6 $\mu$m \citep{Huang2022}, but only of 36.4~au in the 1.3~mm continuum, with an inclination of 27.6 $\pm$ 0.3\deg\ and a position angle of 170.0 $\pm$ 0.9\deg\ \citep{Long2019}. In the CO molecular lines, \citet{Braun2021} reported a disk radius of 202 au. DO Tau is the source of a dense CO outflow whose blueshifted component exhibits multiple rings on a scale of several hundred astronomical units and an associated dynamical timescale of a few hundred years \citep{Fernandez2020}. The system additionally features a bipolar asymmetric optical jet extending about 170 au from the central star at a position angle of 260~$\pm$~3\deg, i.e., perpendicular to the position angle of the disk detected in the millimetric range \citep{Erkal2021}. The jet signature is also seen in a number of strongly blueshifted forbidden emission lines in the optical and near-infrared spectrum of the source (e.g., [OI], [SII], and [FeII]), with blueshifted peak velocities reaching up to about -100 \kms\ \citep{Nisini2024}. \cite{Erkal2021} reported that the jet and counter-jet asymmetry and velocities have remained stable over at least 8 years, carrying a mass flux of about 5~10$^{-9}$\msunyr\, and originating from an magnetohydrodynamic (MHD) disk wind launched at a distance of less than 0.5~au from the central star. 

The inner disk region has been investigated both indirectly through the analysis of specific spectral diagnostics, and directly through long baseline interferometry. \cite{Kidder2021} derived a dust temperature of 1447~$\pm$~110 K at the inner disk edge by fitting the veiling in the H and K bands in the spectral energy distribution. \citet{Alcala2021} estimated the location of the dust sublimation radius at \rsub\ = 0.04~$\pm$~0.01~au from their stellar and accretion luminosities, and adopting for the dust sublimation temperature the effective temperature they derived by fitting the veiling variation as a function of the wavelength with a black-body power law. From the two-telescope K-band Keck Interferometer, \cite{Eisner2014} derived an inner dust radius \rdust\ of 0.18~$\pm$~0.01~au by fitting their two measurements with a uniform ring model. The authors retrieved a size of  \rbrg\ = 0.09~$\pm$~0.01~au for the \brg\ emitting region. From the 2.3 $\mu$m CO bandheads in emission, they also deduced a maximum size \rco\ $\leq$~0.04~au for the gaseous inner disk edge, which agrees with the CO radius estimate, \rco\ = 0.03~au, derived by \cite{Salyk2011} from the width of the 4.65--5.10 $\mu$m rovibrational CO bandheads at a temperature of 1575~$\pm$~200~K.

   \begin{figure*}[t]
   \centering
   \includegraphics[width=0.27\hsize]{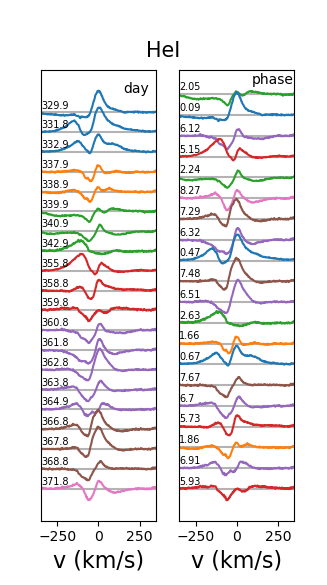}
      \includegraphics[width=0.27\hsize]{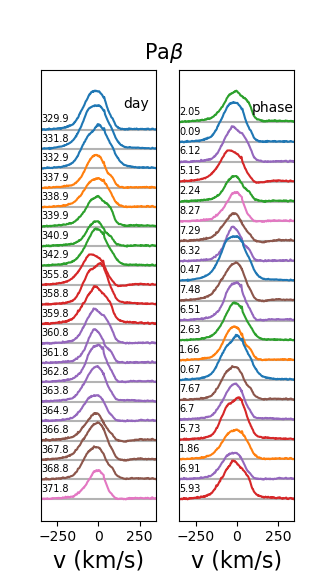}
      \includegraphics[width=0.27\hsize]{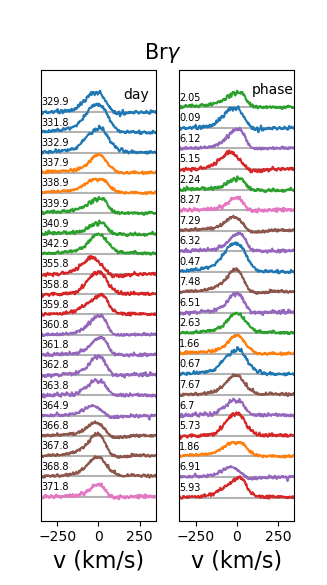}

   \includegraphics[width=0.27\hsize]{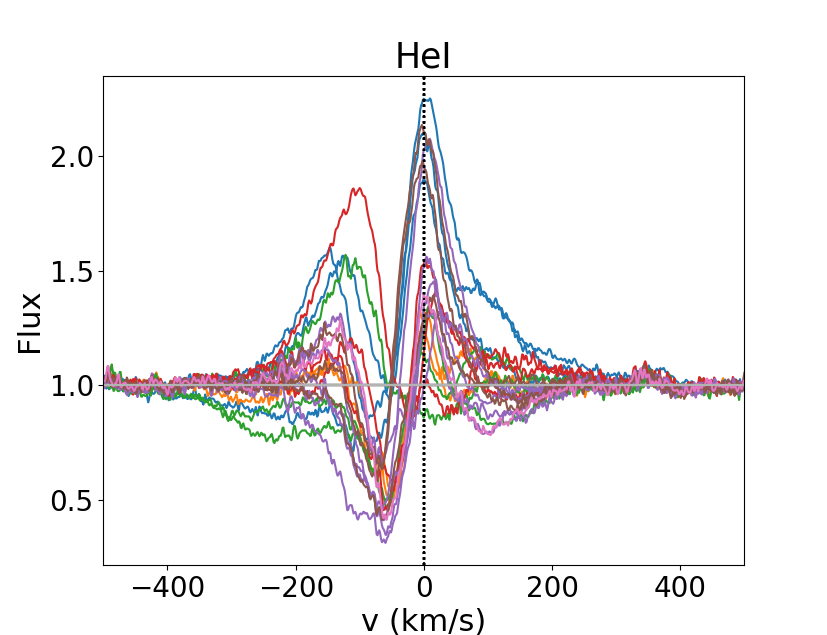}
      \includegraphics[width=0.27\hsize]{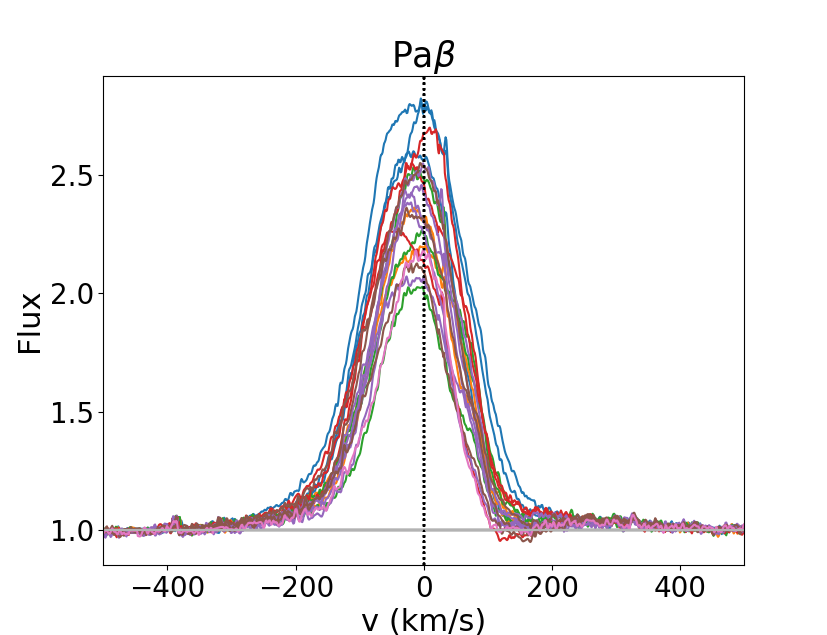}
      \includegraphics[width=0.27\hsize]{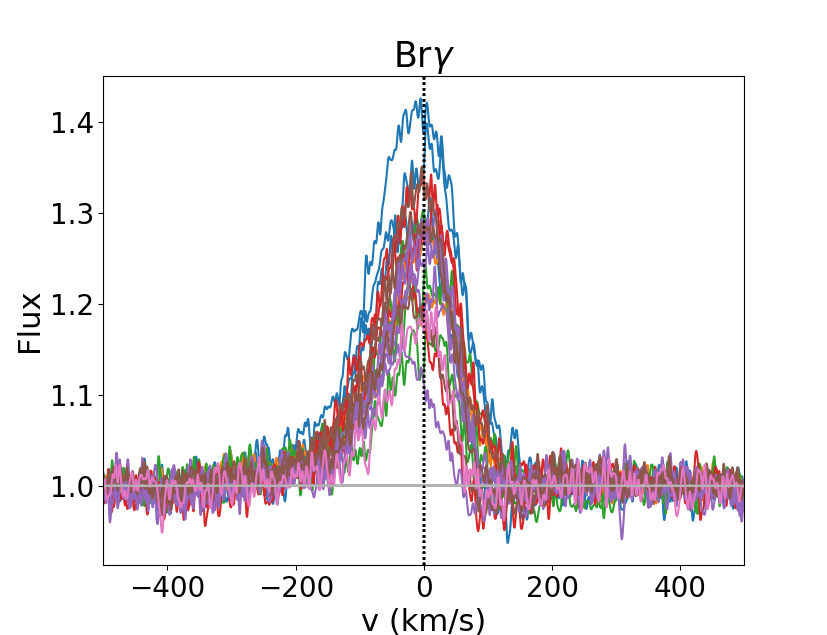}

   \includegraphics[width=0.27\hsize]{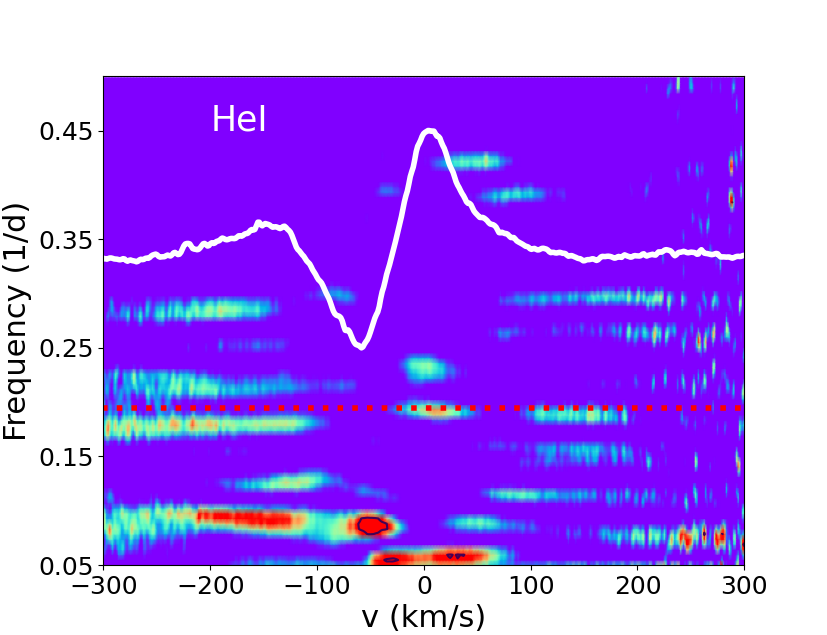}
     \includegraphics[width=0.27\hsize]{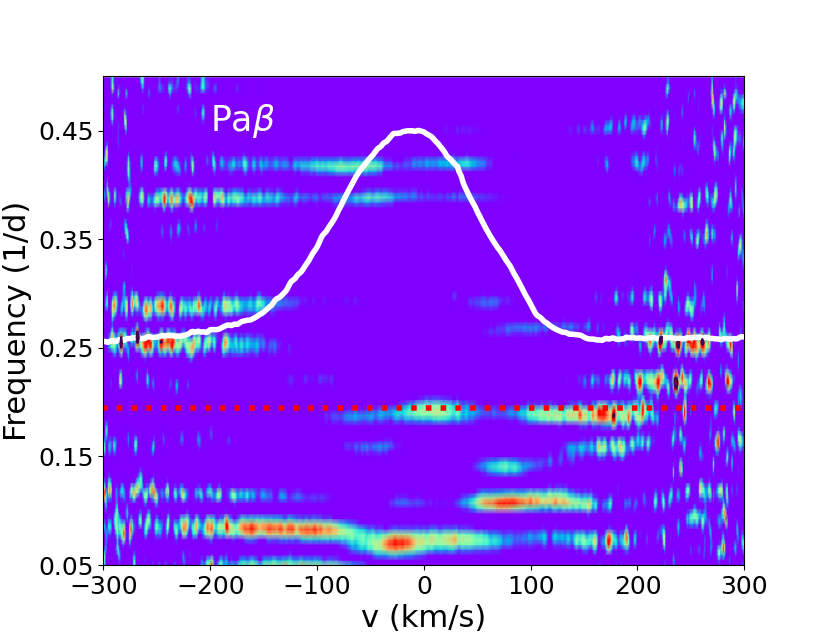}
      \includegraphics[width=0.27\hsize]{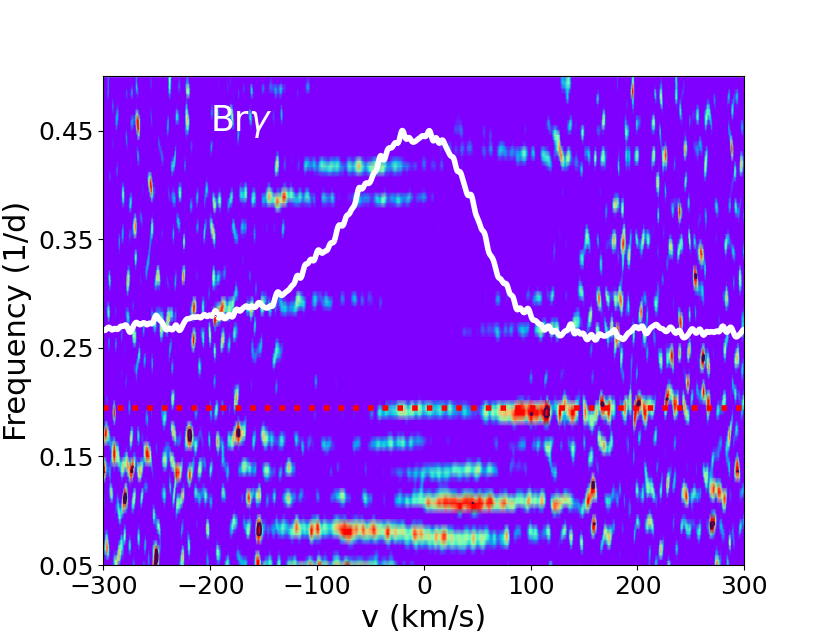}
   \caption{Top: Series of the near-infrared line profiles \hei\ 1083 nm, \pab, and \brg\ plotted as a function of JD-2,460,000 ({left}) and rotational phase ({right}). The color code corresponds to successive rotational cycles. Middle: Plotted profiles superimposed in a single image to illustrate their variability. The vertical dotted line indicates the stellar rest velocity. Bottom: Two-dimensional periodograms across the line profiles. The color code reflects the periodogram power, ranging from 20 ({blue}) to 60\% ({red}) of the maximum power. {The FAP level of 0.1 is shown as black contours.} The horizontal dotted red line drawn at a frequency of 0.195 day$^{-1}$ indicates the star's rotational period (P~=~5.128 d). The white curve is the mean line profile.}
              \label{fig:lineprof}%
    \end{figure*}
   
\section{Near-infrared spectroscopy}
\label{sect:spirou}

\subsection{Observations}

We observed DO Tau with SPIRou \citep{Donati2020}, the near-infrared spectropolarimeter mounted on the 3.6-m CFHT atop Maunakea in Hawaii. SPIRou collects unpolarized and polarized stellar spectra over the 0.95–2.50 $\mu$m wavelength range at a resolving power of 70,000 in a single exposure. We obtained 20 spectra between January 19 and March 1, 2024, with an exposure time of 420~s {each} that yielded a signal-to-noise ratio around 150 in the H band. All spectra were processed with the 0.7.288 version of the SPIRou data reduction software APERO, which also includes telluric correction \citep{Cook2022}. The spectra were normalized to the continuum level. The barycentric stellar radial velocity and veiling in the YJHK bands were computed using a weak-line T Tauri star (TWA 25) as a template of similar {spectral type \citep[M0.5,][]{Herczeg2014}}, following the method described in \cite{Sousa2023}. {In brief, the template spectrum  {\citep[\vrad = 8.33 $\pm$ 0.25 \kms,][]{Nicholson2021}} is shifted to the observed spectrum of DO Tau over several spectral windows that include unblended photospheric lines and a continuum flux is added until DO Tau's spectrum is correctly fit through a $\chi^2$ method. The wavelength shift measures the radial velocity difference between DO Tau's and the mean radial velocity of the template, while the added continuum flux measures the veiling in specific wavelength ranges.} The veiling $r$ measures the ratio between the excess continuum flux ($f_{excess}$) and the photospheric flux ($f_{star}$), with the former arising from either the accretion shock (mostly in the UV and optical ranges) or the dust thermal emission (at infrared wavelengths). Its determination allows us to estimate the contribution of the stellar flux to the total flux in the K band (see Sect. 4.2). The template was veiled, rotationally broadened, and {shifted in radial velocity} to match the spectrum of DO Tau, allowing us to subtract the photospheric lines and compute the residual emission line profiles. The barycentric stellar radial velocity and the veiling in the YJHK bands are given in Table~\ref{tab:vradveiling}. 

\subsection {Results}

The main emission lines seen in DO Tau's near-infrared spectrum are HeI~$\lambda$1083, and HI \pab\ and \brg. Table~\ref{tab:vradveiling} lists their equivalent widths that were measured by direct integration under the residual line profiles {and were not corrected for veiling.} Additional emission features include the forbidden [FeII] 1257.02 and 1644.00 nm lines and the H$_2$ 2121.83 nm line, with the former species blueshifted by -105~\kms and the latter by -17~\kms, therefore presumably originating from the jet and outflow (see Fig.~\ref{fig:feiih2}).

Figure~\ref{fig:lineprof} shows the variability of the HeI~$\lambda$1083, \pab, and \brg\ line profiles as a function of date and rotational phase\footnote{The ephemeris for the rotational phase is computed with \prot=5.128 d and an origin of phase on JD 2,460,329.412, just before the first SPIRou spectrum. This ensures that the fractional phases used in this paper are the same as in the accompanying paper \citep{Donati2026}.} during the 2024A SPIRou runs.  The HeI line profile exhibits strong variability in both blue and red wings, as well as in its peak intensity. Both blueshifted and redshifted absorptions are clearly seen, extending to nearly -400 and +250 \kms, respectively, indicative of simultaneous outflow and accretion processes. In contrast, the \pab\ and \brg\ profiles consist mainly in a single emission component, whose peak intensity varies on a timescale of days. While the profiles are asymmetric with a narrower red wing and a more extended blue wing, only marginal redshifted absorptions can be seen below the continuum on some dates over the velocity range 100-150 \kms. 

A periodogram analysis of the line profiles is shown in the bottom panels of  Fig.~\ref{fig:lineprof}. In the red wing of the \brg\ profile, {significant  power appears at velocities around +100 \kms\ at a frequency corresponding to the stellar rotational period of 5.128~d, with a False Alarm Probability (FAP) peaking at 0.1.} This suggests the periodic crossing of corotating funnel flows on the line of sight. {While some power is seen at the same location in the \pab\ line profile, none of it reaches an FAP of 0.1.} In the HeI line profile, the strongest power suggests a longer period of 12 days and is located at negative velocities around -50 \kms, corresponding to the deepest blueshifted absorption component in the line profile. This signature appears to extend down to $-200$~\kms\ across the blueshifted wing of the HeI line profile, {albeit at a lower significance level (FAP $>$ 0.1)}. {The periodicity seen in the blueshifted wing of the HeI profile may result from a non-axisymmetric outflow originating from a region located beyond the disk corotation radius. Finally, the HeI profile  also exhibits some power around zero velocity, extending to $\pm$~40~\kms, at a period of about 17  days, peaking at FAP of 0.1. Given the limited time span of our observations, namely 42 days interrupted by a 2-week gap, we believe this longer period needs to be confirmed by more extended observations.}

Figure~\ref{fig:ewveiling}-top shows the YJHK veiling values as a function of Julian date during the 2024A SPIRou runs. While the veiling is relatively mild in the YJH bands, it increases significantly in the K~band. The veiling values in the YJH bands and their weak variations are similar to those previously reported for this source by \cite{Sousa2023} from SPIRou spectra obtained in 2020. Yet, the veiling in the K~band has slightly increased compared to these previous epochs, from an average r$_k$=2.33 in 2020 to r$_K$=3.0-4.0 in 2024. Figure~\ref{fig:ewveiling}-bottom shows the variations of the HeI~$\lambda$1083, \pab, and \brg\ line equivalent widths. Little change has occurred in the equivalent widths of the three lines compared to previous observations in 2020 \citep{Sousa2023}. The \pab\ and \brg\ line equivalent widths vary by a factor of 2 to 4 on a timescale of days, and we note that the largest line equivalent widths correspond to maximum veiling values, both peaking on JD 2,460,331.805. Indeed, Fig.~\ref{fig:ewvsveiling} reveals a clear correlation between the line equivalent widths $EW$ and the nearby continuum veiling $r$, implying significant line flux variations, since the line flux $F_{line}$ is proportional to $(1+r)~\times~EW$. {A periodogram analysis of the veiling-corrected \pab\ and \brg\ line equivalent widths does not provide any convincing results. Yet when the veiling-corrected EW's are plotted as a function of rotational phase, as shown in Fig.~\ref{fig:ewveilcor}, and regardless of the peak emission occurring at phase 0.47 that likely corresponds to an accretion burst on JD 2,460,331.805 (see above), there is marginal evidence of modulation along the rotational cycle. Interestingly, the minimum line flux occurs near phase 0.3, when the redshifted absorptions are the most conspicuous in the HeI line profile (see Fig.~\ref{fig:lineprof}), which suggests that the line emitting region is partially hidden from view as the accretion funnel flow crosses the line of sight.}

   \begin{figure}[t]
   \centering
   \includegraphics[width=0.85\hsize]{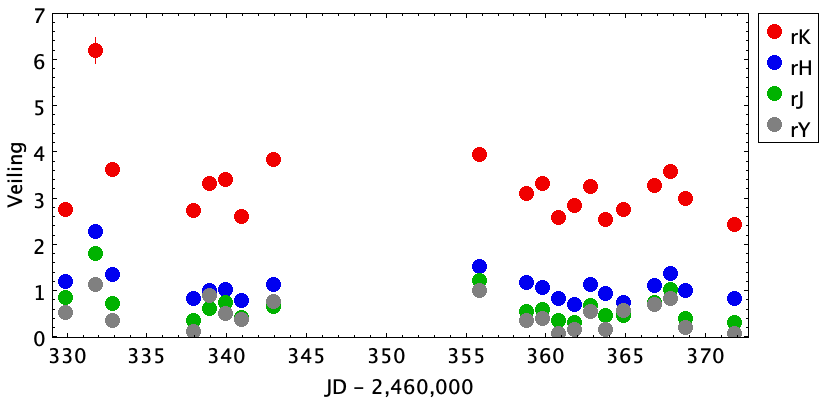}
      \includegraphics[width=0.85\hsize]{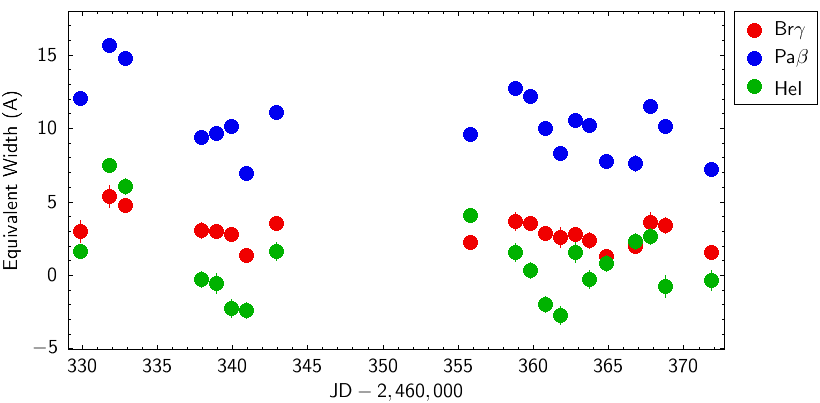}
   \caption{Near-infrared veiling in the YJHK bands ({\it top}) and line equivalent widths for the HeI 1083 nm, \pab, and \brg\ line profiles ({\it bottom}) as a function of Julian date. The measurement error is usually smaller than the symbol size (see Table~\ref{tab:vradveiling}).}
              \label{fig:ewveiling}%
    \end{figure} 

   \begin{figure}[t]
   \centering
   \includegraphics[width=0.7\hsize]{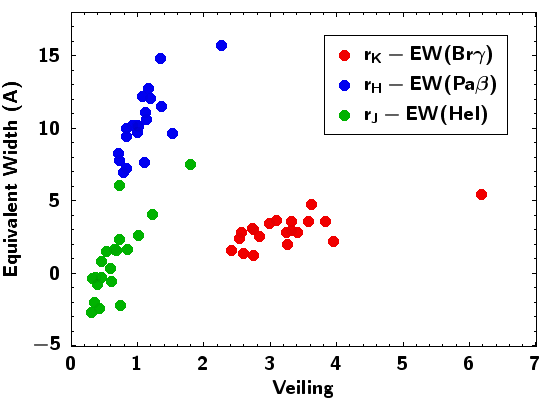}
   \caption{Equivalent width of the \hei, \pab, and BrG\ lines as a function of veiling in the JHK bands. Each line is compared to the veiling within the same wavelength range. }
              \label{fig:ewvsveiling}%
    \end{figure} 

   \begin{figure}[t]
   \centering
   \includegraphics[width=0.9\hsize]{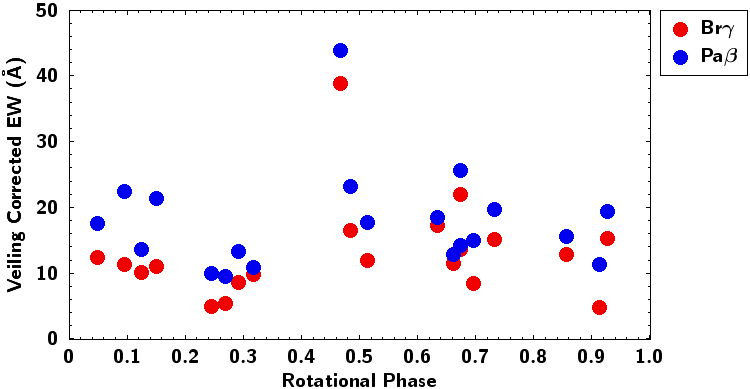}
   \caption{{Veiling corrected equivalent widths, EW$\times$(1+r),  of the \pab\ and \brg\ lines as a function of rotational phase.}}
              \label{fig:ewveilcor}%
    \end{figure} 
\subsection{Mass accretion rate}

We derived the mass accretion rate from the flux of the near-infrared lines \pab\ and \brg, which minimizes the impact of the large uncertainties affecting the visual extinction on the line of sight reported in the literature (A$_V$~=~0.75–-3.04 mag; see Sect. 2). We computed the line flux from the line equivalent widths measured on SPIRou spectra, EW(\pab)~=~7--16 \AA\  and EW(\brg)~=~1.3--5.4 \AA, and since we lacked contemporaneous near-infrared photometry, we adopted the J-band and K-band 2MASS magnitudes (J~=~9.47, K~=~7.3) to estimate the nearby continuum flux\footnote{We could not find more recent K-band magnitude measurements in the literature. However, \cite{Park2021} provide 13 measurements of WISE W1 (3.6 $\mu$m) photometry for DO Tau obtained between Feb. 2014 and Apr. 2020. The mean value is W1 = 6.10 mag and the standard deviation is 0.16 mag, which suggests that DO Tau's near-infrared flux does not drastically vary over the years.}. Assuming a distance of 138.5 pc and propagating the uncertainties on A$_V$\footnote{The large range of $A_V$ from 0.75 to 3.04 mag results in a factor of 1.8 uncertainty in L(\pab) and of 1.3 in L(\brg).}, we derived the line luminosities L(\pab)~=~0.9--3.9~10$^{30}$~erg.s$^{-1}$ and L(\brg)~=~1.6--8.1~10$^{29}$~erg.s$^{-1}$ for the extreme values of the line equivalent widths. Using the \lacc - \lline\ relationships of \cite{alcala2017}, we derived accretion luminosities ranging from 2.1~10$^{32}$ to 2.4~10$^{33}$~erg.s$^{-1}$ (0.06--0.4 \lsun). Alternatively, using the \lacc-\lline\ relationships revisited by \cite{Fiorellino2025}, we obtained \lacc = 8.4 10$^{32}$--7.0~10$^{33}$~erg.s$^{-1}$ (0.22--1.8~\lsun). 

To compute the mass accretion rate from Eq. (1) of \cite{alcala2017}, we adopted the dynamical mass \mstar~=~0.54~\msun\ and the radius we derived from the rotational period (\rstar~=~1.8~$R_\odot$; see Sect. 2), as this determination is more accurate than the determination based on the luminosity, as the latter is affected by the uncertainty on extinction. Assuming a truncation radius of 2.5~\rstar\ \citep{Pittman2025}, we obtained \macc~=~9.6~10$^{-9}$--1.1~10$^{-7}$~\msunyr and \macc~=~3.810$^{-8}$--3.2~10$^{-7}$~\msunyr\, when using the \lacc\ range derived above from the \lacc - \lline\ relationships of \cite{alcala2017} and \cite{Fiorellino2025}, respectively. The range of accretion rates thus spans more than an order of magnitude, mostly due to intrinsically varying line fluxes, and is also partly impacted by systematics depending on which \lacc-\lline\ relationships were selected. 

These estimates are consistent with those of previous studies (see Sect. 2) and confirm that DO Tau is a strong accretor, with a mass accretion rate about 10 times larger than the mean accretion rate measured for Class II T Tauri stars of the same mass \citep[see][]{Manara2023}.  

\section{Interferometry}

\subsection{GRAVITY observations and data reduction}

We observed DO~Tau with the four Auxiliary Telescopes (ATs) and the four Unit Telescopes (UTs) of the VLTI on December 24, 2022, and on December 28, 2023, respectively, as part of the Guarantee Time Observations of the GRAVITY consortium. We used {GRAVITY in its single field mode, with the fringe tracker \citep[FT;][]{Lacour2019} locking the fringes on DO~Tau itself} at a frame rate of $\sim$300~Hz and of $\sim$900~Hz for the ATs and the UTs, respectively. We used integration times as long as 30~s on the scientific detector of GRAVITY. We recorded 3 files of 6-min each with the medium configuration of the ATs (telescopes D0, G2, J3, and K0), and 10 files of 6-min each with the UTs. All files are recorded with the high spectral resolution mode (R $\sim$ 4000). These sequences on DO~Tau were interleaved with observations on interferometric calibrators to properly calibrate the instrumental transfer function. For the ATs, the weather conditions were excellent with a seeing ranging between 0.52" and 0.68", and a coherence time in the K~band up to 39~ms, which allows the GRAVITY fringe tracker to perfectly lock the fringes despite the K-band magnitude of 7.3, which was close to the limiting magnitude with the ATs at that time. For the UTs, the seeing during the observations ranged between 0.48" and 0.77", and the coherence time in the K~band was about 7-8~ms. The log of the interferometric observations is given in Table~\ref{tab:log}. The GRAVITY observations with the UTs were obtained about three weeks before the SPIRou spectropolarimetric observations presented above. 

We used the GRAVITY pipeline \citep{DRS} to reduce and calibrate the observations. We obtained K-band continuum squared visibilities and closure phases for probing the inner rim of the protoplanetary disk (Sect. 4.2), and differential visibilities and phases across the HI \brg~line for studying the accretion-ejection processes in the star-disk interaction region. Despite the excellent weather conditions, the signal-to-noise ratio of the AT observations was not high enough to obtain measurable differential visibilities and phases across the HI \brg~line. We thus only used the higher-quality UT data for modeling the \brg~line emitting region (Sect. 4.3).

\begin{table*}[t]
    \begin{center}
    \caption{Log of the GRAVITY observations of DO~Tau. }
    \begin{tabular}{l l l l l l l l}
    \hline
    \hline
    HJD & Date & Time & Configuration & N & Seeing & $\tau_0$ & Calibrators\\
    &  & (UT) &  &  & ('') & (ms) & \\
    \hline
2459937.614583 & 2022-12-24 & 02:31-02:59 & D0-G2-J3-K0 & {3} & 0.52-0.68 & 22.6-39.1 & HD 33589\\
 & & & & & & & 2MASSJ04333368+23374\\ [1ex]
2460306.671528 & 2023-12-28 & 03:29-04:44 & UT1-2-3-4 & 10 & 0.48-0.77 & 6.6-8.4 & HD 31464\\
    \hline
    \end{tabular}
    \end{center}
    \noindent {\bf Notes.} The columns provide the date and the time of the observations, the telescope configuration, the number of exposures (N), the seeing and the coherence time $\tau_0$ of the observations, and the calibrators.
    \label{tab:log}
\end{table*}

\begin{figure*}[t]
  \centering
  \includegraphics[width=0.58\hsize]{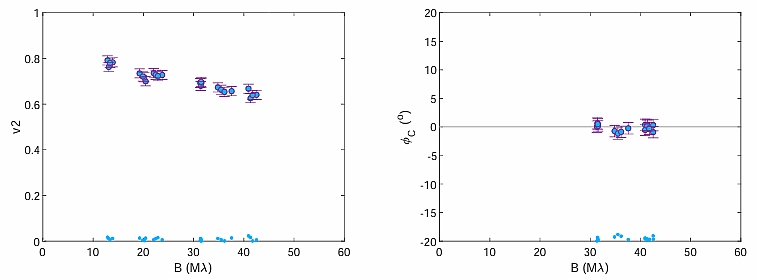}  \includegraphics[width=0.22\hsize]{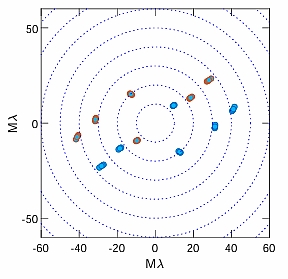}
  \includegraphics[width=0.8\hsize]{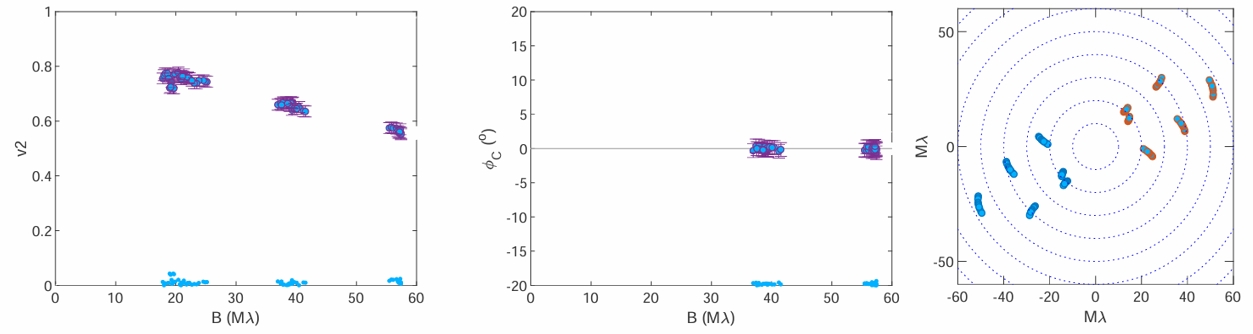} 
  \caption{GRAVITY observations of DO~Tau with the AT (top) and with the UT (bottom) configurations of the VLTI. The visibilities squared (left) and closure phases (middle) for the central spectral channel of binned data ($\lambda$~=~2.15~$\mu$m) are plotted as a function of the spatial frequency. The corresponding {\it (u, v)} planes over the K band are plotted in the right panels. The blue symbols at the bottom of the visibility and closure phase curves display the amplitude of the residuals of the best-fit ring models. See text for detail. 
  }
  \label{fig:GRAVITY_AT}%
\end{figure*}
   
\subsection{Modeling the K-band continuum emission}

In the near-infrared, the continuum emission is dominated by the dust located close to the dust sublimation front. To determine the properties of this K-band emission, {as the error bars provided by the GRAVITY reduction pipeline can be underestimated
and/or correlated,} we {conservatively} applied 2\% floor values to the error estimates on the squared visibilities, and used the geometrical model proposed by \citet{Lazareff2017} as previously used in \citet{gravity_2019,gravity_2021}. It includes three components: a point-like source for the central source, an inner disk modeled as a Gaussian ring, and an extended component, the so-called halo, which is assumed to be fully resolved. As explained in \citet{Lazareff2017}, when the inner disk is only partially resolved, its flux contribution and its size are degenerate (see their Sect. 3.4 and their Fig. 4). We therefore used the veiling values $r_K$ derived from SPIRou measurements (see Sect.~\ref{sect:spirou}) to fix the starting value for the stellar flux contribution $f_s$ and kept it as a free parameter in our fitting process. For $r_K$ ranging between 2.5 and 4 (Fig.~\ref{fig:ewveiling}), the stellar flux contribution ranges between 20\% and 30\%. We fixed the spectral index\footnote{The spectral index is defined as $k~=~\frac{d \log F_\nu}{d \log \nu}$, with $F$ the flux and $\nu$ the frequency.} of the star $k_s$ from the effective temperature of the star using Table 5 of \citet{pecaut_2013}. We are left with seven free parameters: the flux contributions of the extended component $f_{halo}$ and of the inner disk $f_d$ (with $f_s + f_d + f_{halo}$~=~1), the spectral index $k_d$, the half-flux radius $a_K$, the inclination $i$, the position angle $PA$, and the width-to-half-flux ratio $w$ of the inner disk. The corresponding model for the visibility as a function of spatial frequencies {\it (u, v)} and wavelength $\lambda$ is expressed by
\begin{equation}
    V (u,v,\lambda) = \frac{f_s (\lambda_0/\lambda)^{\rm k_s} + f_d (\lambda_0/\lambda)^{\rm k_d} V_d (u,v)}{(f_s + f_{halo}) (\lambda_0/\lambda)^{\rm k_s} + f_d (\lambda_0/\lambda)^{\rm k_d}},
\end{equation}
with $\lambda_0$ the central wavelength of the K band, and $V_d$ the visibility of a pure Gaussian ring to model the inner disk rim {(see Appendix C)}; the visibility of the star equals to 1 as it is a point-like source, and that of the halo is null as it is fully resolved. 

{We then fit the interferometric quantities by adjusting the free parameters of our model using the same tools as those previously used by \cite{Lazareff2017} and \cite{gravity_2019,gravity_2021}. The $\chi^2$ minimization is done through two steps of the Shuffled Complex Evolution algorithm and one final MCMC process, as detailed in Sect. 3.3 of \cite{Lazareff2017}.}

To derive the inclination $i$ and the position angle $PA$ of the inner disk, we first focused on the AT data only and binned the scientific data exhibiting 5 spectral channels over the K band, as the corresponding {\it (u, v)} plane contains two perpendicular short baselines (see Fig.~\ref{fig:GRAVITY_AT}-top-right). {The circumstellar environment of} DO~Tau appears to be partially resolved with the AT baselines, with squared visibilities ranging from 0.6 to 0.8. We detect no clear departure from centro-symmetry as the closure phases are consistent with 0. When fitting the squared visibilities with our model, we obtained an inclination $i$ of 49$\pm$5$^\circ$ and a $PA$ of 166$\pm$5$^\circ$, which corresponds to an almost north-south direction that is unfortunately not well probed by our (u, v) plane. {We obtained a flux contribution for the halo of 11\%, and an half-flux radius of 0.7~mas for a star flux contribution of 15\%. To refine the size of the environment, we then fixed these values of inclination and position angle,} and fit together the AT and UT datasets (Fig.~\ref{fig:GRAVITY_AT}), as the visibilities squared for the ATs and the UTs are similar at the shorter baselines (i.e., $\sim$~20~M$\lambda$) despite the different fields of view (i.e., $\sim$~250~mas and $\sim$~60~mas for the ATs and the UTs, respectively). Our best-fit corresponds to a $\chi^2$ of 0.8, maybe due to a too large floor values of the error estimates. For the best ring model, we obtained a flux contribution of $f_{d}$~=~66\%, an half-flux radius $a_K$ of 0.62$\pm$0.03~mas (i.e., 0.086$\pm$0.004~au at 138.5~pc), and a width-to-half flux ratio $w$ consistent with 1, which could be due to our limited angular resolution that prevents us from resolving the inner cavity (if there is one). Regarding the extended component, we derived a flux contribution of $f_{halo}$~=~12\%, while the similar values of the visibilities at the shorter baselines suggest a size on the order of the field of view of the UTs. All the best-model parameters are given in Table~\ref{tab:fitGRAVITY}. 

\subsection{Modeling the HI Br$\gamma$ emitting region}

\begin{figure*}[t]
  \centering
  \includegraphics[width=1\hsize]{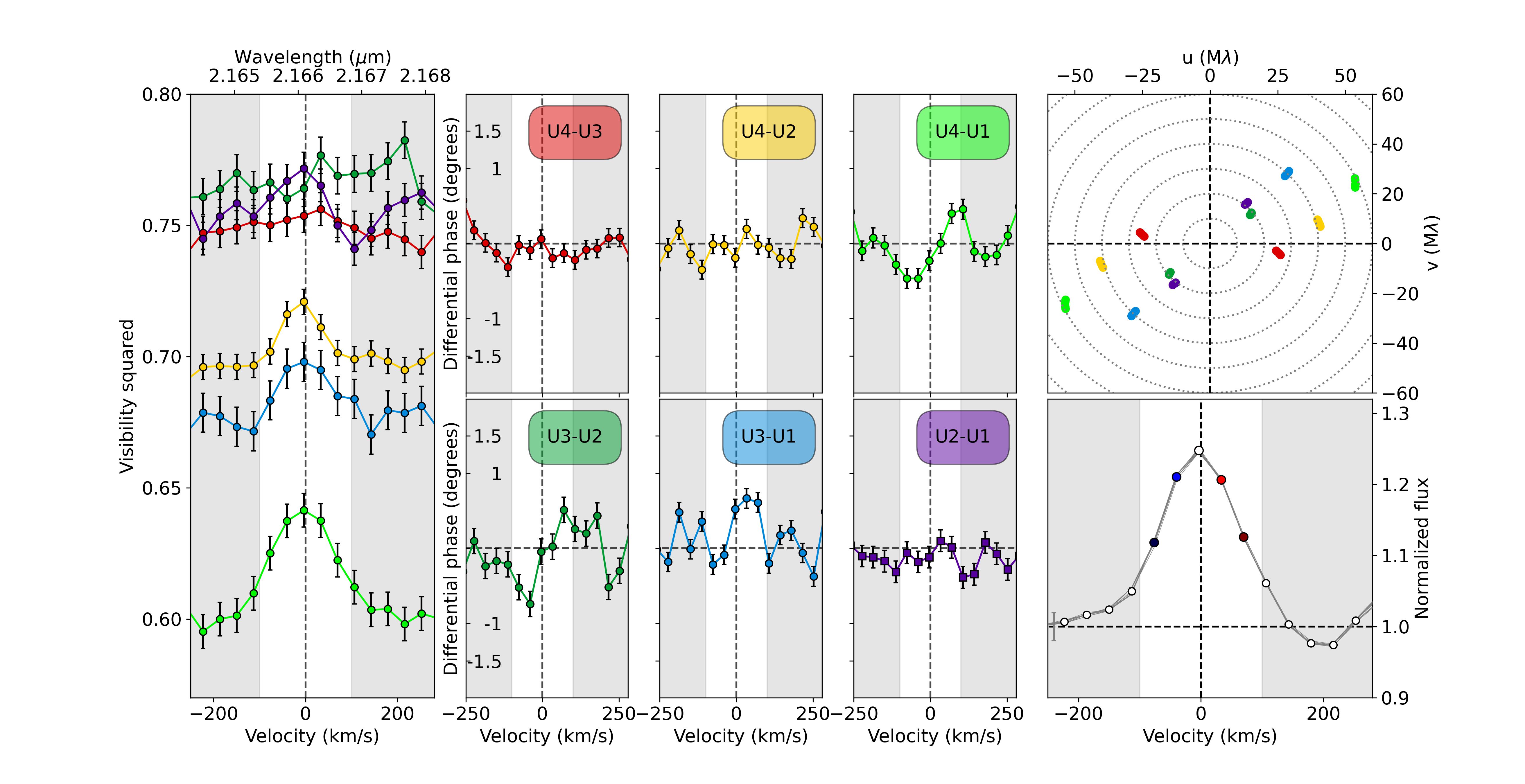}
  \caption{\brg{} differential interferometric observables of DO~Tau: Squared differential visibilities (left) and differential phases (middle) for the six color coded interferometric baselines depicted in the {\it (u, v)} plane (upper-right); line-to-continuum flux ratio in the reference frame of the star, LSR- and tellurics-corrected (lower-right). For the interferometric observables, the plots give the median value over the whole observing sequence. Gray shaded regions correspond to the velocities that are not considered for our \brg\ line analysis.}
  \label{fig:diff_data}%
\end{figure*}

\begin{figure*}[t]
  \centering
  \includegraphics[width=0.9\hsize]{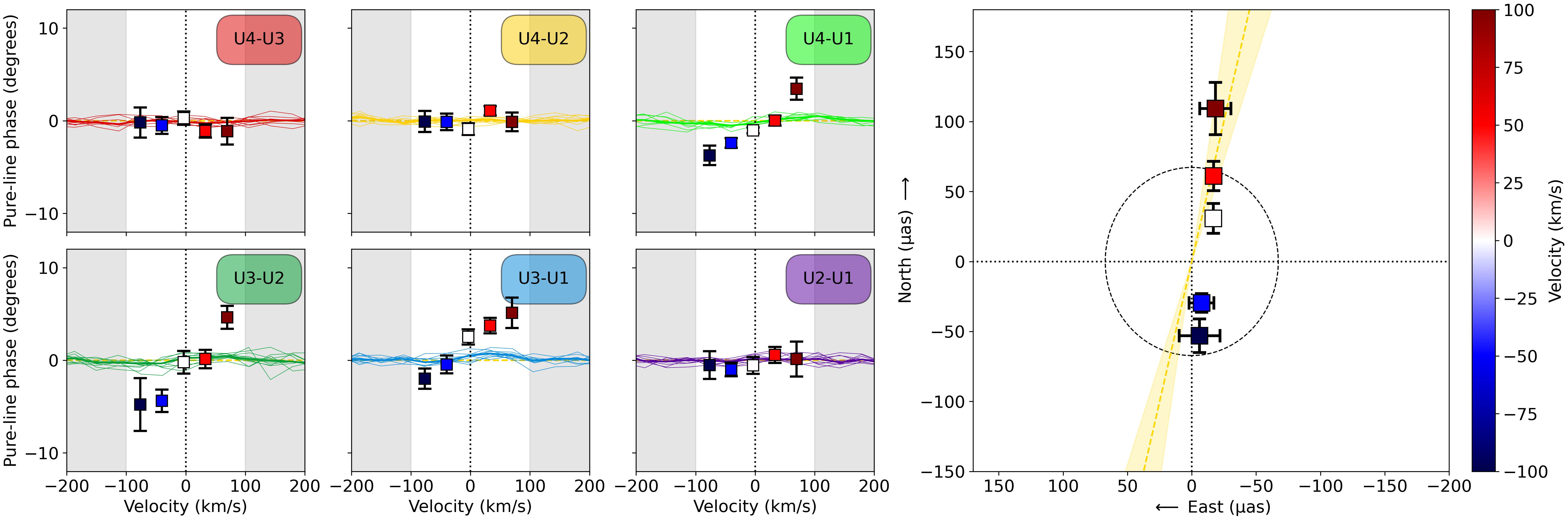}
  \caption{\brg{} pure-line differential phases (left) and on-sky photocenter shifts (right) with respect to the continuum photocenter (at (0, 0)) as a function of velocity coded from blue to red. On the pure-line phases, the colored lines correspond to the observed total differential phases. On the photocenter plot, the light-yellow line and cones correspond to the position angle of the inner disk and its uncertainty, respectively. The dotted black circle represents the size of the star.}
  \label{fig:photocenters}%
\end{figure*}

Thanks to the spectral resolution of GRAVITY and the high signal-to-noise ratio reached with the UTs, we were able to observe the HI \brg{} line in the K-band spectrum of DO~Tau, and to record differential interferometric observables across this emission line (Fig.~\ref{fig:diff_data}). The \brg{} line exhibits a peak line-to-continuum ratio of 1.25, a high-velocity blueshifted wing, and a slight absorption in the red wing that is consistent with accretion (bottom-right panel). For the longest interferometric baselines (U1-U4, U2-U4, and U1-U3), the visibilities across the line are clearly larger than the neighboring continuum visibilities (left panel), which indicates that the \brg{} line emitting region is more compact than the continuum one. A non-null differential phase signal across the emission line is detected for three baselines (U2-U3, U1-U3 and U1-U4; middle panel). An S-shape profile is observed for the differential phases on U1-U4 and marginally on U2-U3, which generally traces a rotation process \citep[see for example][]{Mendigutia2015}.

\begin{table}[t]
        \caption{Best-fit parameters and 1$\sigma$ errors for the K-band and \brg\ line emitting regions.}
    \begin{center}
    \begin{tabular}{l l}
    \hline
    \hline
    Parameters & Values  \\
    \hline
    {\bf K-band emission} &  \\  [1ex]
    Fraction of star flux $f_s$ & $ 22 \pm 5$\%  \\  [1ex]   
    Fraction of halo flux $f_h$ & $ 12 \pm 5$\%  \\ [1ex]
    Fraction of ring flux $f_c$ & $ 66 \pm 5$\%    \\ [1ex]
    Inclination $i$ & $ 49^\circ \pm 5^\circ$   \\ [1ex]
    Position Angle $PA$ & $ 166^\circ \pm 5^\circ$  \\ [1ex]
    Half-flux radius $a_K$ &  $ 0.62\pm 0.03$ mas   \\ [1ex]
    Half-flux radius $R_K$ & $ 0.086 \pm 0.004$ au \\ [1ex]
    Width-to-half-flux ratio $w$ & $ 0.98^{+0.02}_{-0.15}$ \\ [1ex]
    Spectral index $k_d$ & $-1.4 \pm 0.3$  \\ [1ex]
    \hline
    $\chi^2_r$ & 0.8  \\ 
    \hline
    {\bf Br$\gamma$ emission} &\\  [1ex]
    Fraction of halo flux $f_h$ & $ 10 \pm 2$\%  \\ [1ex]
    Half-Width Half-Maximum$^{(*)}$ & $ 0.08 \pm 0.05$~mas \\ [1ex]
     & $ 0.011 \pm 0.007$~au\\ [1ex]
    \hline
    $\chi^2_r$ & 0.6  \\   
    \hline\hline
\end{tabular}
\end{center}
\noindent {\bf Notes.} $^{(*)}$ for the central spectral channel.
\label{tab:fitGRAVITY}
\end{table}

To study the kinematics of the emitting region only, we extracted the pure-line differential visibilities $V_{\rm line}$ and phases $\phi_{\rm line}$ using the same method as described in \citet{wojtczak_2023}, while the errors are obtained through a classical propagation. Since the \brg{} emission is only partially resolved, the on-sky displacements $\bold{p}$ of each spectral channel with respect to the photocenter of the continuum could be directly obtained from the pure-line phases \citep{Lachaume2003}:
\begin{equation}
    \bold{p}.\bold{B} = -\lambda\,\frac{\phi_{\rm line}}{2\pi},
\end{equation}
where $\bold{B}$ are the projected baselines and $\lambda$ is the wavelength of a given spectral channel. 
The pure-line phases and the corresponding on-sky displacements are depicted in Fig.\ref{fig:photocenters}. The latter draws a line $\sim$200~$\mu$as (i.e., $\sim$0.03~au) long aligned along the position angle of the inner disk (depicted in yellow). The red and blue channels depart from the zero-velocity channel (white symbol) in the northwest and southeast direction, respectively. 

We modeled the pure-line visibilities with the same model as described in \citet{nowacki_2024}:
\begin{equation}
    V_{\rm line} = (1-C_h)\,\exp{\left(-\frac{\pi^2 B_{\rm eff}^2 \Theta^2}{\lambda^2 \ln{2}}\right)},
\end{equation}
where $C_h$ is the fraction of the total flux attributed to a fully resolved halo component, $\Theta$ is the half-width at half-maximum (HWHM) of a Gaussian disk, and $B_{\rm eff}$ is the effective projected baseline which accounts for a potential inclination of the Gaussian disk. This model was adjusted to the \brg{} pure-line visibilities in each spectral channel using independent MCMC procedures for each spectral channel running  in parallel, where the priors are Gaussian distributions with central values and dispersions set to the continuum best values and associated uncertainties. We used 500 walkers over 2500 steps in order to probe the models' parameters and reach convergence. Because the \brg{} pure-line emission is only partially resolved, there is a degeneracy between $C_h$ and $\Theta$. Therefore, we first adjusted the size of the disk alone, while fixing $C_h$~=~12\% as in the continuum (see Sect. {4.2}). Once this first convergence was reached, both parameters were let free until a second convergence was reached. We therefore provided a measurement of the typical size for the \brg{} emitting region, given that this component is only marginally resolved with GRAVITY. The errors on the best parameters were estimated thanks to the posterior distribution of each parameter (see Fig. D.3 for the corner plot). We obtained a size of $R_{Br\gamma}$~=~0.011~$\pm$~0.007~au ($\sim$~1.4~\rstar) in the core of the line and a halo contribution similar to that derived for the continuum ($10~\pm~2$ \%), which is consistent with an origin from light scattering \citep{Pinte2008}.
All the parameters of the best model and their uncertainties are listed in Table~\ref{tab:fitGRAVITY} (see Fig~\ref{fig:V2curve} for a visibility curve of the best model superimposed to the data).

\begin{figure*}[t]
  \centering
  \includegraphics[width=0.62\hsize]{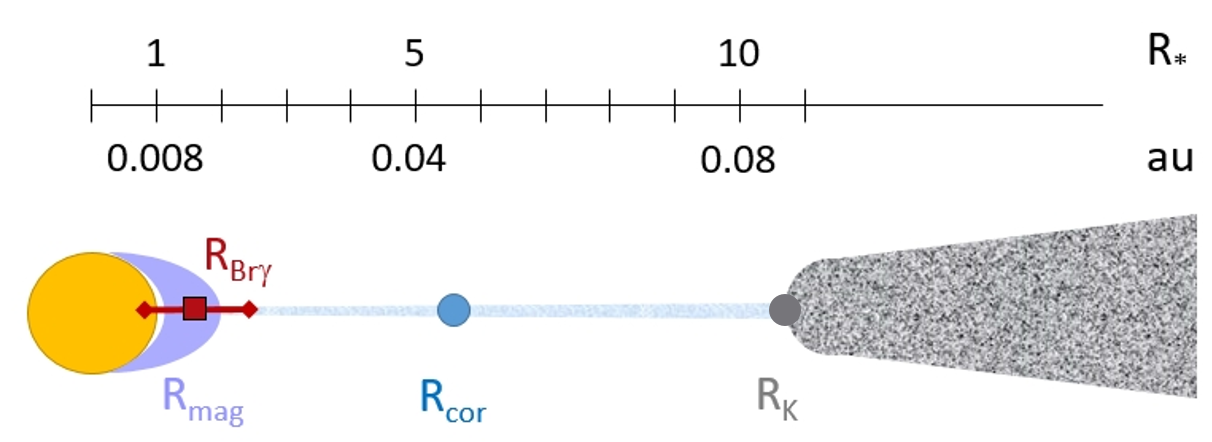}
  \caption{Characteristic sizes for DO~Tau: HWHM of \brg{} emitting region $R_{Br\gamma}$ (red square), magnetospheric truncation radius $R_{\rm mag}$ (purple area), corotation radius $R_{\rm cor}$ (blue circle), and half-flux radius of the K-band continuum emission $R_K$ (gray).
  }
  \label{fig:diff_size}
\end{figure*}

\section{Discussion}

We have spectrally and spatially resolved the star-disk interaction region of DO Tau thanks to near-infrared spectra provided by SPIRou at CFHT, and K-band interferometric measurements with GRAVITY at VLTI. Based on the complementary results brought by these two techniques, we discuss in the following sections the main properties related to the accretion/ejection processes in this young, low-mass star as it interacts with its strongly accreting circumstellar disk.

\subsection{The accretion regime}

For this low-mass (0.54~\msun) T Tauri star, we derived a mass accretion rate in the range $\sim$10$^{-8}$--10$^{-7}$~\msunyr, classifying DO~Tau as a relatively strong accretor. When looking at the line flux of the \pab\ and \brg\ lines over the sequence of observations, we found evidence of a periodic modulation in the redshifted, high velocity wings of the profiles at the rotational period of the star of 5.128 d. This suggests the presence of large-scale accretion funnel flows persistent over at least seven rotational periods of the system. The HI and HeI near-infrared line profiles exhibit significant variability on a timescale of days. The latter has deep absorption features seen in both the blue and the red wings, with no clear sign of phase consistency from one rotational period to the next. This suggests significant variations of the mass accretion rate onto the star on short timescales, as also indicated by the burster-type K2 light curve the source displays \citep{Cody2022}. This finding is supported by the low ratio between the magnetic truncation radius and the corotation radius. Using the rotational period and stellar mass listed in Table~\ref{tab:param}, we computed a  corotation radius, i.e., the location where the Keplerian velocity of the disk equals the stellar angular velocity, $R_{\rm cor}$~$\sim$~0.047$\pm$0.002~au~$\sim$~5.6~\rstar. To compute the magnetospheric truncation radius $R_{\rm mag}$, where the stellar magnetic field truncates the inner disk, we used the formula from \citet{hartmann2016}:
\begin{equation}
    R_{\rm mag} = 12.6 \frac{B_d^{4/7}{R_*}^{12/7}}{{M_*}^{1/7}\dot{M}_{\rm acc}^{2/7}},
\end{equation}
where the dipolar magnetic field contribution at the equator $B_d$ is scaled to 1~kG, the stellar radius $R_*$ to 2~$R_\odot$, the stellar mass $M_*$ to 0.5~$M_\odot$, and the mass accretion rate $\dot{M}_{\rm acc}$ to 10$^{-8}$~\msunyr. With the parameters of Table~\ref{tab:param}, we obtained a $R_{\rm mag}$ range of 0.008--0.016~au ($\sim$~1-2~\rstar) for a dipolar magnetic field strength $B_d$ of 0.3~kG derived from the SPIRou observations \citep{Donati2026}. 
The upper limit of $R_{\rm mag}$ does not exceed 2~\rstar, leading to a ratio $R_{\rm mag}$/$R_{\rm cor}$ below 0.34, indicative of an  ordered unstable accretion regime \citep{Blinova2016}. According to the simulations presented by these authors, in this regime, chaotic unstable accretion tongues merge to form one or two "ordered" accretion tongues that penetrate into the magnetosphere where they transition into regular funnel flows. In the case of a large inclination of the magnetic axis relative to the stellar rotational axis, as was the case for DO~Tau in 2024 \citep[$\beta \sim 50^\circ$;][] {Donati2026}, \cite{Kulkarni2008, Kulkarni2009} showed that the instability is reduced because the magnetic pole is closer to the disk plane, making it easier to load the accreting material onto the magnetic field lines. Then, a modulation at the stellar rotation period is expected and this would account for the rotational modulation of the accretion flow we detected in the high-velocity redshifted wings of the line profiles at the stellar rotation period (see Sect.~3.2).

\subsection{The origin of the Br$\gamma$ emission line}

From the GRAVITY measurements, we derived a very compact size for the \brg{} line emitting region  ($R_{Br\gamma}$~=~0.011~$\pm$~0.007~au~$\sim$~1.4~\rstar) of the same order as the magnetospheric truncation radius (see Fig.~\ref{fig:diff_size}). While our simple geometrical modeling of the \brg\ interferometric visibilities may underestimate the actual size of the magnetospheric cavity by up to a factor 2 compared to radiative transfer models of magnetospheric \brg\ emission \citep{Tessore2023}, the corotation radius ($R_{\rm cor}$ $\sim$~5.6~\rstar) remains much larger. This suggests that the \brg\ emission is dominated by magnetospheric accretion and/or an inner wind at the magnetosphere-disk interface \citep[e.g.,][]{Najita2000}. When comparing with the modeling of \cite{Wojtczak2024}, the constant \brg{} size across blue to red spectral channels of the emission line (see Appendix D) favors their pure magnetospheric accretion scenario, as a disk wind contribution increases the size in the line wings \citep[see] [for the example of the strong accretor S~CrA~North]{nowacki_2024}. Nevertheless, \cite{Wojtczak2024} only considered a disk wind (starting at 7-9~\rstar) and did not model a wind launched very close to the star. In addition, the SPIRou He~I line profiles exhibit a clear blueshifted absorption (see first column of Fig. 1) that is either narrow or large (up to -300 km/s), which suggests the presence of both a disk wind and a stellar and/or inner wind \citep{Kwan2007}. This is corroborated by previous spectroscopic observations of DO~Tau \citep{Alencar2000,Banzatti2019,Nisini2024} that are well modeled by a conical-shell inner wind \citep{Kurosawa2012}. The \brg{} emission of DO~Tau might thus partially include a contribution of outflowing material very close to the stellar surface.

The on-sky photocenter shifts across the \brg{} emission line provided by GRAVITY lie along the position angle of the disk (Fig.~\ref{fig:photocenters}). As detailed in \citet{Tessore2023}, these displacements can trace the accretion flow along the rotation of the star and of the magnetosphere. GRAVITY observations were acquired at a rotational phase of the system corresponding to 0.77 in \citet{Tessore2023}'s modeling\footnote{In \citet{Tessore2023}, phase 0.5 refers to the epoch when the main accretion column faces the observer. This corresponds to phase $\sim$0.3 of our ephemeris, when the deepest redshifted absorptions are seen in the line profiles (Fig.~\ref{fig:lineprof}). The GRAVITY observations were obtained at phase 0.57 according to our ephemeris, which corresponds to a rotational phase of $\sim$0.77 in \citet{Tessore2023}'s definition.}. We can use their 3D radiative transfer simulations computed at a rotational phase of 0.69 (see their Fig. 5), closest in phase to the GRAVITY observations, as they considered an inclination of 60$^\circ$, close to that of DO~Tau ($\sim$~49$^\circ$). Our on-sky displacements (orientation and amplitude) exhibit a similar slight misalignment with the disk's position angle as predicted from their radiative transfer model at this rotational phase. However, we cannot completely rule out another scenario (e.g., a stellar and/or inner wind), as the obliquity of the magnetic field of DO~Tau \citep[$\sim$~50$^\circ$;][] {Donati2026} is higher than what is considered in their simulations (obliquity of 10$^\circ$).

\subsection{The inner rim of the disk}

In the near-infrared range, the continuum emission is dominated by the rim of the inner disk where the dust is sublimated by the irradiation of the central star. We first estimated the expected sublimation radius $R_{\rm sub}$ (in au) for DO~Tau from the relationship provided by \cite{Monnier2002}:
\begin{equation}
    R_{\rm sub} = 1.1 \sqrt{Q_R} \sqrt{\frac{L_*}{1000 L_\odot}} \left( \frac{1500}{T_{\rm sub}} \right)^2,
\end{equation}
which yields $R_{\rm sub}$ ranging between 0.02 and 0.06~au for a stellar luminosity of 0.44~$\pm$~0.08~$L_\odot$, a typical sublimation temperature of silicates $T_{\rm sub}$ between 1300~K and 1700~K, and a ratio of the absorption efficiencies of the dust of the incident field and of the reemitted field $Q_R$~=~3 for an effective temperature of about 3500~K \citep[see Fig. 2 of][] {Monnier2002}. Although roughly similar within uncertainties, this sublimation radius (0.02 -- 0.06~au) is 1.4-4 times smaller than the half-flux radius determined from GRAVITY observations (0.086~$\pm$~0.004~au; {see Sect. 4.2}). We then developed radiative transfer models with MCMax \citep{Min2009} for different dust populations and found that a single population of small (0.1~$\mu$m) olivine grains or a mix of small (0.1~$\mu$m) olivine and iron grains provide inner rim positions consistent with the GRAVITY half-flux radius measurement (see Appendix~\ref{app:disk} for a detailed description of the model). Alternatively, since DO~Tau is a strong accretor, the inner rim could include larger grains and be driven further out due to hot accretion spot(s) on the stellar surface and/or viscous heating: when taking into account the accretion luminosity \lacc{} (Sect. 3.3) in Eq.~(5) in addition to the stellar luminosity, we obtained sublimation radii as large as 0.1~au, consistent with GRAVITY's continuum measurement. \citet{Eisner2014} derived a size about twice as large, \rsub = 0.18~$\pm$~0.01~au, which may be due to the fact that they neglected the halo contribution, thus overestimated the continuum ring size \citep{Pinte2008}. Alternatively, DO~Tau might have been in a different accretion regime at the time of their observations with enhanced heating processes, as the authors determined a \brg\  equivalent width of 11.9~\AA, i.e., nearly three times as large as the values we derived here.

\subsection{A misalignment of the outer disk}

From the GRAVITY data, we determined an inclination for the inner disk of 45-55$^\circ$, which is larger than the inclination of the outer disk as derived with ALMA \citep[{27.6$^\circ$;}][]{Long2019}. A higher inclination for the star is also more consistent with the magnetic field map reconstructed from Zeeman-Doppler imaging \citep{Donati2026}. It is also more in agreement with its fundamental parameters, as using the ALMA disk's inclination for the star would lead to a stellar radius of 2.8~\rsun, which disagrees with the tracks from \cite{Feiden2016} for a star with the mass, gravity, and effective temperature of DO~Tau, as listed in Table~\ref{tab:param}. As previously observed for LkCa~15 \citep{Alencar2018}, RX~J1604.3-2130A \citep{Sicilia-Aguilar2020}, and DoAr~44 \citep{Bouvier2020b}, the inclinations of the central star and of the inner part of the disk appear to be similar, while the outer disk is misaligned. 

Although warps and misalignments of protoplanetary disks are common \citep{Marino2015,Benisty2017,Bohn2022}, the origin of such morphologies is still a matter of debate. Among the mechanisms that could generate a warp of the disk, an interaction between the disk and a tilted magnetosphere has been invoked to explain the light curve of AA Tau \citep{Bouvier1999, Terquem2000}. In the grid of 3D magnetohydrodynamic simulations performed by \citet{Romanova2021}, different angles of misalignment (5-10$^\circ$, and up to 30-40$^\circ$) between the stellar rotation axis and the normal of the disk could be obtained depending on the initial position of the inner disk, the rotational period, and the magnetic moment of the star. 

Another possibility to generate a warp is the presence of a companion on an inclined orbit \citep{Nealon2020}. This scenario has been investigated by \citet{Erkal2021} to account for the slight jet wiggling they detected in their observations in the [Fe~II] line at 1644~nm, as wiggling is generally expected to trace the dynamical impact of an unseen companion. As the faster, redshifted jet exhibits a larger wiggling amplitude than the blueshifted one, the authors ruled out the "orbital scenario" \citep{Masciadri2002}, where the wiggling originates from the reflex motion of the jet source under the influence of the companion. \citet{Erkal2021} also investigated the "precession scenario" \citep{Terquem2019}, where the jet wiggling could trace the disk precession due to tidal interactions with a companion in a non-coplanar orbit, and found that it is not fully in agreement with the observations either. They thus favored a precession induced by a warping instability driven by magnetic torques, which agrees well with the variability of the magnetic field topology of DO~Tau over month- to year-timescales detected with SPIRou observations \citep{Donati2026}. 

\citet{Dullemond2019} and \citet{Kuffmeier2021} proposed that the misalignment of the outer disk could be the result of late infall events during which matter is accreted with a misaligned angular momentum. This scenario sounds plausible, when considering the complex environment of DO~Tau with arc-like structures and the suspected past stellar encounter with the neighboring HV~Tau system \citep{Winter2018}. 

Accurately locating the disk's warp would aid toward understanding its origin. Unfortunately, this is not reachable with the existing datasets, since the protoplanetary disk of DO~Tau is probed at the $\sim$~0.1~au scale with GRAVITY and at a few tens of astronomical units with ALMA. Complementing these datasets with MATISSE observations on intermediate scales would potentially provide a full sampling of the disk structure. 

\section{Conclusions}

By combining spectroscopic and K-band interferometric observations of the T Tauri star DO Tau, we spatially and spectrally resolved the central astronomical unit of the system, and this allowed us to investigate the physical processes at play in the star-disk interaction region. Our main conclusions are as follows:
\begin{itemize}
    \item With a mass accretion rate of 10$^{-8}$--10$^{-7}$~\msunyr, {we confirm that} this low-mass ($\sim$~0.5~\msun) star {is} a strong accretor. {We found a periodic modulation of the high-velocity redshifted wing of the HI \brg\ line profile} at the stellar rotational period, which suggests the existence of corotating accretion funnel flows linking the inner disk to the stellar surface. We derived a ratio $R_{\rm mag}$/$R_{\rm cor}$ of 0.34 or less, which points toward an ordered unstable accretion regime, as expected for a star classified as a burster.
    \item  The size we measured for the Br$\gamma$ line emitting region is very compact ($\sim$~0.01~au). It is much smaller than the size of the K-band continuum emitting region ($\sim$~0.09~au) and on the order of the magnetospheric truncation radius. This suggests a dominant contribution of magnetospheric accretion and/or a wind originating very close to the star for the Br$\gamma$ emission.
    \item The on-sky displacements we measured across the \brg\ emission line profile are almost aligned along the PA of the disk on a scale of $\sim$~200~$\mu$-arcseconds (i.e., $\sim$~0.03~au), and they possibly trace a corotating magnetospheric accretion flow. 
    \item We derived an inclination for the inner disk that agrees with the star's inclination, and it is larger by 15-25$^\circ$ than that of the outer disk. This points to a misalignment of the outer disk that could be related to the past encounter between DO~Tau and the close by HV~Tau system.
\end{itemize}

The results reported here demonstrate that the magnetospheric accretion process extends down to low-mass strongly accreting T Tauri stars, such as DO~Tau. They illustrate the power of combining {contemporaneous (and ideally simultaneous)} high-resolution spectroscopic and long baseline interferometric observations  to probe the physical processes at play at the star-disk interaction region of young stellar systems. The companion paper \citep{Donati2026} provides even further constraints by deriving, through  spectropolarimetric analysis, the magnetic strength and topology of the large-scale stellar magnetic field responsible for this interaction.

\section*{Data availability}
SPIRou data used in this study are, or will soon be, publicly available at the Canadian Astronomy Data Center (\url{https://www.cadc-ccda.hia-iha.nrc-cnrc.gc.ca/}). The GRAVITY data are publicly available in the ESO archive (\url{https://archive.eso.org/}). 

\begin{acknowledgements}
We thank the anonymous referee for the fruitful suggestions that improved the paper. We thank George Pantolmos for enlightening discussions on the unstable accretion regime in young stars and Myriam Benisty for fruitful discussions.
This work has been supported by the French National Research Agency (ANR) in the framework of the “Investissements d’Avenir” program (ANR-15-IDEX-02) and in the framework of the “ANR-23-EDIR0001-01” project (IRYSS project). This research has made use of the Jean-Marie Mariotti Center Aspro \url{(http://www.jmmc.fr/aspro)} and SearchCal \url{(http://www.jmmc.fr/searchcal)} services co-developed
 by LAGRANGE and IPAG, and of CDS Astronomical Databases SIMBAD and VIZIER \url{(http://cdsweb.u-strasbg.fr/)}. This work has made use of data from the European Space Agency (ESA) mission {\it Gaia} (\url{https://www.cosmos.esa.int/gaia}), processed by the {\it Gaia} Data Processing and Analysis Consortium (DPAC, \url{https://www.cosmos.esa.int/web/gaia/dpac/consortium}). Funding for the DPAC has been provided by national institutions, in particular the institutions participating in the {\it Gaia} Multilateral Agreement. A.C.G. acknowledges the support from PRIN-MUR 2022 20228JPA3A “The path to star and planet formation in the JWST era (PATH)” funded by NextGeneration EU and by INAF-GoG 2022 “NIR-dark Accretion Outbursts in Massive Young stellar objects (NAOMY)” and Large Gran INAF-2024 “Spectral Key features of Young stellar objects: Wind-Accretion LinKs Explored in the infraRed (SKYWALKER)”. A.K. acknowledges support by the NKFIH NKKP grant ADVANCED 149943 and the NKFIH excellence grant TKP2021-NKTA-64. Project no.149943 has been implemented with the support provided by the Ministry of Culture and Innovation of Hungary from the National Research, Development and Innovation Fund, financed under the NKKP ADVANCED funding scheme. A.A and P.G. acknowledge support from Fundaç\~ao para a Ciência e Tecnologia (FCT), Portugal,  through grants UID/99/2025 and  UID/50007/2025. L.L. and E.B. gratefully acknowledge the Collaborative Research Center 1601 funded by the Deutsche Forschungsgemeinschaft (DFG, German Research Foundation)—SFB 1601 [sub-project A3].
 
\end{acknowledgements}\bibliographystyle{aa}
\bibliography{Biblio}

\begin{appendix}

\section{SPIRou spectra: Radial velocity, veiling, and equivalent width measurements}
\label{app:spirou}

Table~\ref{tab:vradveiling} lists the radial velocity, veiling in the YJHK bands, and equivalent width for the \hei, \pab, and \brg\ lines. These measurements are extracted for each SPIRou spectrum. 

\section{Fe[II] and H2 lines in the SPIRou spectra}
\label{app:feiih2}

Fig.~\ref{fig:feiih2} shows the Fe[II] 1257.02 and 1644.00 nm, and the H$_2$ 2121.83 nm line profiles from the SPIRou spectra. For each line, all the spectra are superimposed to monitor the variability over the observing sequence.

\section{Ring model for the K-band emission}

{To model the K-band continuum emission from the inner rim where dust grains sublimate, we convolve a wireframe image with an ellipsoidal kernel with a Gaussian radial distribution, as described in detail in Sect. 3.6 of \cite{Lazareff2017}. The wireframe distribution has a complex visibility given by (their Eq.~7)
\begin{equation}
   V_{wire}(q) = J_0(2\pi~q~a_r),
\end{equation}
and the Gaussian kernel has a complex visibility given by (see their Table 5)
\begin{equation}
   V_{Gauss}(q) = \exp{\left(-\frac{(\pi a_{G}~q)^2}{\ln{2}}\right)},
\end{equation}
with $a_r$ the ring angular radius, and $a_{G}$ the kernel angular radius.
The resulting complex visibility is
\begin{equation}
   V_d(q) = V_{wire}(q) . V_{Gauss}(q)
.\end{equation}
The physical quantities (see their Table 7) are the half-flux radius $a_K=\sqrt{a_{G}^2 + a_r^2}$ and the width-to-half-flux ratio $w~=~a_{G}/a_{K}$ .
}

\section{Br$\gamma$ emission best-fit model}
\label{app:brg}

Once extracted, the pure-line visibilities are modeled by Eq. (3) given in Sect. 4.3. The best-fit model and the corresponding residuals are displayed in Fig.~\ref{fig:V2curve}. The sizes derived for each velocity are given in Table~\ref{tab:hwhmbrg} and illustrated in Fig.~\ref{fig:Gsizes}. {We give the corner plot of the associated MCMC process in Fig.~\ref{fig:Gcorner}.}

\begin{figure}[t]
    \centering
    \includegraphics[width=\linewidth]{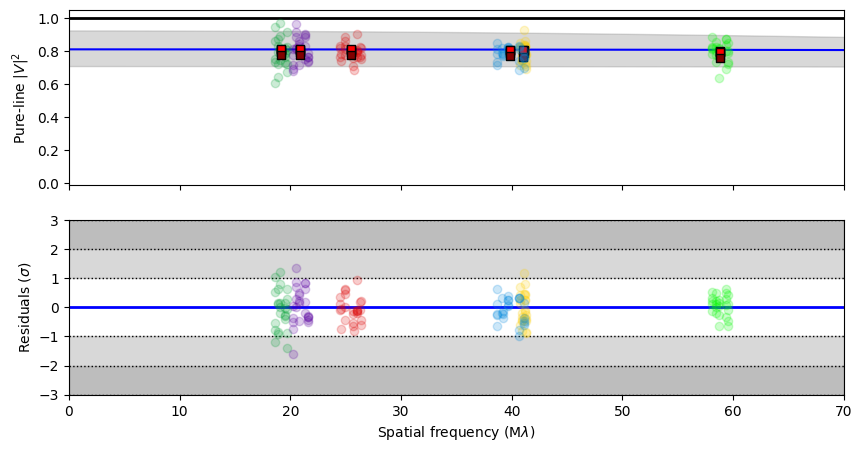}
    \caption{Top: Pure-line visibility squared as a function of the spatial frequency: GRAVITY measurements for the six baselines (colored symbols as in Fig. 4) and best-fit model (continuous blue line), surrounded by its 3$\sigma$ uncertainty (gray shaded area). Bottom: Residuals of best-fit model expressed in units of uncertainties of the data.}
    \label{fig:V2curve}
\end{figure}

\begin{figure}[h]
    \centering
    \includegraphics[width=0.75\linewidth]{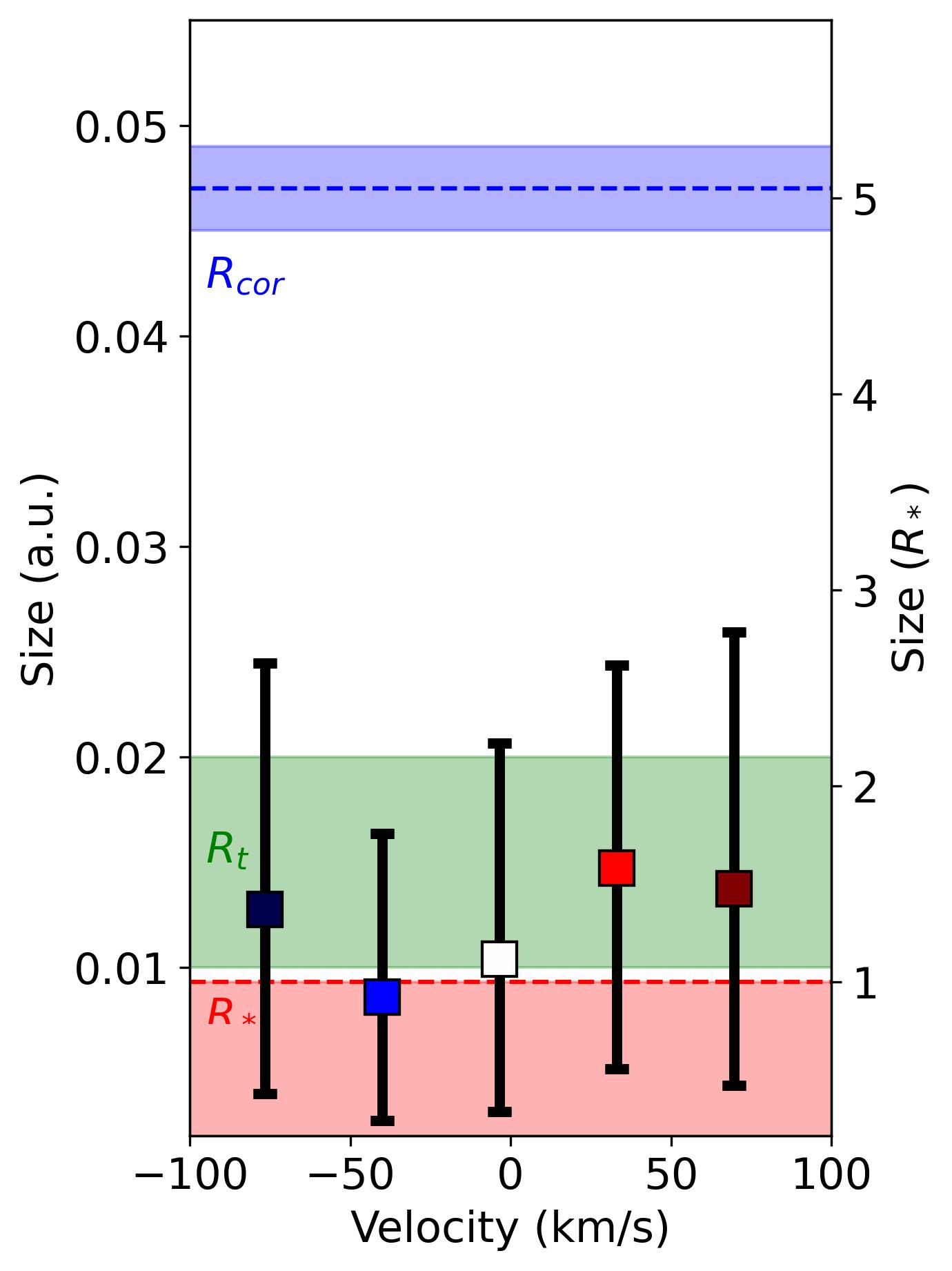}
    \caption{Half-width at half-maximum of the best Gaussian disk model for the \brg{} line emitting region as a function of the spectral channel. For comparison, the red, green and blue areas illustrate the stellar, the magnetospheric truncation, and the corotation radius, respectively. }
    \label{fig:Gsizes}
\end{figure}

\begin{table}[b]
\centering
\caption{Half-width at half-maximum and contribution of halo ($C_h$) of the \brg\ emission for each velocity channel with 1$\sigma$ errors.}
\label{tab:hwhmbrg}
\begin{tabular}{r|cc}
\hline \hline
Velocity [km/s]   & HWHM [{mas}] & $C_h$ [\%]        \\
\hline
-77 & $0.09^{+0.09}_{-0.06}$ & $11 \pm 2$ \\ [1ex]
-40 & $0.06^{+0.06}_{-0.04}$ & $10.1^{+0.9}_{-1.0}$ \\ [1ex]
-4 & $0.08 \pm 0.05$ & $10 \pm 1$ \\ [1ex]
32 & $0.10 \pm 0.07$ & $9.7^{+0.7}_{-0.8}$ \\ [1ex] 
69 & $0.10^{+0.09}_{-0.07}$ & $12^{+1}_{-2}$ \\ [1ex] 
\hline
\end{tabular}
\end{table}

\begin{figure*}[h]
    \centering
    \includegraphics[width=\linewidth]{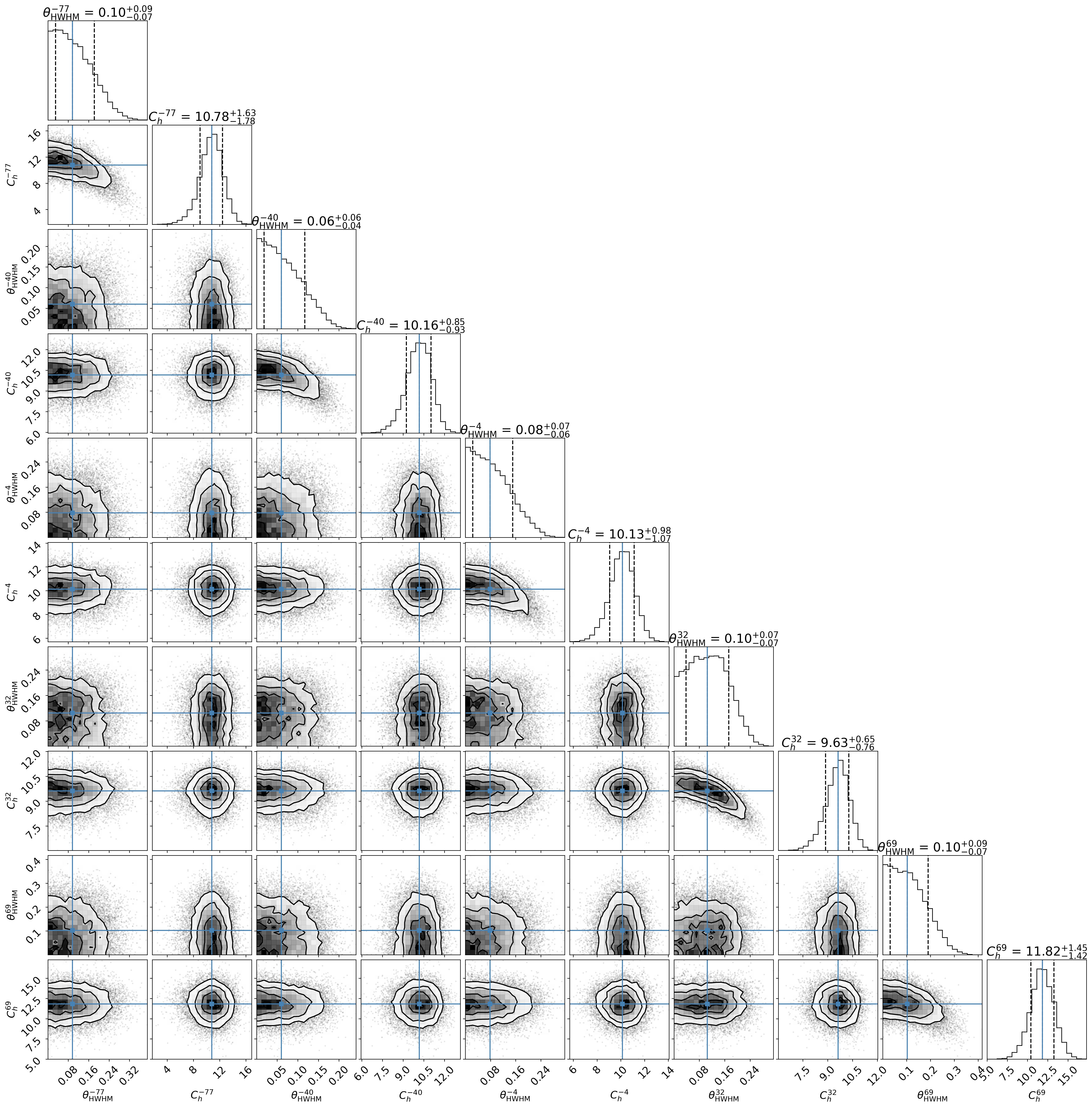}
    \caption{{Corner plot associated with the MCMC process performed on pure-line quantities. The contribution of the halo and the half-width at half maximum for a given velocity $v$ are denoted as $\theta_{\rm HWHM}^v$ and $C_h^v$, and expressed in percentage and mas, respectively.}}
    \label{fig:Gcorner}
\end{figure*}

\section{Self-consistent disk models}
\label{app:disk}

We used the radiative transfer code MCMax \citep{Min2009} to compute the structure of protoplanetary disks with different dust properties around DO Tau. MCMax offers a self-consistent treatment of dust sublimation: the dust density structure is solved by iterating the radiative transfer, dust sublimation and recondensation, and vertical hydrostatic equilibrium \citep{Kama2009}. In our simulations, the star was modeled based on the measured parameters of DO Tau ($M_\star=0.54$~$M_\odot$, $T_{\rm eff}=3500$~K, $L_\star=0.44$~$L_\odot$, $i$~=~49$^\circ$; see Table~\ref{tab:param}). We used initial disk boundaries from 0.001 to 200 au, and a power law distribution $\Sigma(r) \propto r^{-1.5}$ for the initial radial profile of the dust surface density. The dust mass in the disk was set at 10~$M_\oplus$, close to the mean of $15\pm3$~$M_\oplus$ found around T Tauri stars in the 1--3~Myr Lupus complex \citep{Ansdell2016}. Four different dust compositions were tested:
\begin{enumerate}
    \item a single population (100\%) of small (0.1~µm) olivine grains,
    \item a mix (80\%/20\%) of small (0.1~µm) and large (100~µm) olivine grains, respectively,
    \item a MRN \citep{Mathis1977} distribution of olivine grains (2\% of 0.1~µm, 4\% of 1~µm, 10\% of 10~µm, 24\% of 100~µm, and 60\% of 1~mm),
    \item a mix (99\%/1\%) of small (0.1~µm) olivine and iron grains, respectively.
\end{enumerate}

   \begin{table*}[t]
\caption{Photospheric radial velocity, near-infrared veiling in the YJHK bands and equivalent widths of the HeI 1083 nm, \pab, and \brg\ lines listed  with their 1$\sigma$ error.}             % title of Table
\label{tab:vradveiling}      % is used to refer this table in the text
\centering                          % used for centering table
  \tiny
\begin{tabular}{l  l l | l l l l l l l l | l l l l l l}     \hline\hline                        
\noalign{\smallskip}
Julian date & \multicolumn{2}{c}{\vrad} & \multicolumn{8}{c}{Veiling} & \multicolumn{6}{c}{|EW| (\AA)} \\
-2,460,000 & \kms & err & r$_Y$ & err &  r$_J$ & err & r$_H$ & err & r$_K$ &err & HeI & err & \pab & err & \brg & err\\
\hline                                   
\noalign{\smallskip}
329.898 & 16.73 & 0.25 & 0.54 & 0.04 & 0.86 & 0.03 & 1.20 & 0.02 & 2.75 & 0.07 & 1.65 & 0.51 & 12.07 & 0.25 & 3.01 & 0.77 \\
331.805 & 17.38 & 0.50 & 1.13 & 0.06 & 1.80 & 0.05 & 2.27 & 0.04 & 6.18 & 0.27 & 7.50 & 0.39 & 15.67 & 0.19 & 5.40 & 0.76 \\
332.867 & 16.15 & 0.09 & 0.35 & 0.04 & 0.73 & 0.02 & 1.34 & 0.02 & 3.62 & 0.09 & 6.06 & 0.52 & 14.80 & 0.29 & 4.76 & 0.47 \\
337.934 & 15.77 & 0.33 & 0.12 & 0.03 & 0.37 & 0.01 & 0.83 & 0.01 & 2.73 & 0.07 & -0.27 & 0.57 & 9.40 & 0.29 & 3.06 & 0.55 \\
338.934 & 15.30 & 0.43 & 0.90 & 0.07 & 0.61 & 0.02 & 1.00 & 0.01 & 3.31 & 0.08 & -0.54 & 0.69 & 9.67 & 0.35 & 2.98 & 0.44 \\
339.922 & 16.14 & 0.75 & 0.51 & 0.04 & 0.74 & 0.02 & 1.02 & 0.01 & 3.41 & 0.10 & -2.22 & 0.60 & 10.11 & 0.27 & 2.79 & 0.46 \\
340.922 & 15.94 & 0.79 & 0.38 & 0.04 & 0.42 & 0.02 & 0.79 & 0.01 & 2.60 & 0.07 & -2.41 & 0.53 & 6.96 & 0.25 & 1.39 & 0.38 \\
342.922 & 15.41 & 0.51 & 0.77 & 0.06 & 0.66 & 0.02 & 1.13 & 0.02 & 3.83 & 0.10 & 1.62 & 0.60 & 11.10 & 0.25 & 3.56 & 0.40 \\
355.828 & 17.12 & 0.40 & 1.00 & 0.05 & 1.23 & 0.03 & 1.53 & 0.02 & 3.95 & 0.12 & 4.08 & 0.48 & 9.61 & 0.19 & 2.22 & 0.49 \\
358.812 & 16.21 & 0.64 & 0.35 & 0.04 & 0.54 & 0.02 & 1.17 & 0.02 & 3.10 & 0.08 & 1.54 & 0.59 & 12.74 & 0.28 & 3.67 & 0.63 \\
359.805 & 15.84 & 0.92 & 0.39 & 0.04 & 0.59 & 0.02 & 1.08 & 0.02 & 3.32 & 0.09 & 0.34 & 0.57 & 12.19 & 0.39 & 3.55 & 0.27 \\
360.816 & 16.18 & 0.34 & 0.08 & 0.03 & 0.36 & 0.01 & 0.83 & 0.01 & 2.57 & 0.06 & -1.99 & 0.53 & 10.01 & 0.28 & 2.83 & 0.31 \\
361.812 & 16.43 & 0.34 & 0.16 & 0.03 & 0.31 & 0.01 & 0.71 & 0.01 & 2.84 & 0.07 & -2.71 & 0.60 & 8.28 & 0.31 & 2.57 & 0.67 \\
362.816 & 15.75 & 0.73 & 0.55 & 0.04 & 0.68 & 0.02 & 1.14 & 0.02 & 3.25 & 0.10 & 1.59 & 0.66 & 10.57 & 0.27 & 2.79 & 0.32 \\
363.750 & 15.78 & 0.68 & 0.17 & 0.03 & 0.47 & 0.02 & 0.93 & 0.01 & 2.54 & 0.07 & -0.26 & 0.62 & 10.21 & 0.29 & 2.40 & 0.55 \\
364.859 & 15.65 & 0.13 & 0.56 & 0.04 & 0.46 & 0.02 & 0.74 & 0.01 & 2.75 & 0.06 & 0.81 & 0.57 & 7.76 & 0.44 & 1.26 & 0.22 \\
366.805 & 16.69 & 0.34 & 0.70 & 0.06 & 0.74 & 0.02 & 1.11 & 0.01 & 3.26 & 0.08 & 2.32 & 0.56 & 7.64 & 0.53 & 2.00 & 0.21 \\
367.789 & 16.69 & 0.24 & 0.83 & 0.05 & 1.02 & 0.03 & 1.36 & 0.02 & 3.58 & 0.09 & 2.63 & 0.53 & 11.52 & 0.26 & 3.59 & 0.67 \\
368.766 & 15.91 & 0.44 & 0.21 & 0.03 & 0.40 & 0.02 & 1.00 & 0.01 & 2.99 & 0.07 & -0.75 & 0.76 & 10.17 & 0.35 & 3.41 & 0.52 \\
371.820 & 16.11 & 0.42 & 0.09 & 0.03 & 0.32 & 0.02 & 0.83 & 0.01 & 2.42 & 0.06 & -0.33 & 0.65 & 7.20 & 0.34 & 1.56 & 0.28 \\
\hline                     
\hline                                   
\end{tabular}
\end{table*}

\begin{figure*}[t]
  \centering
  \includegraphics[width=0.33\hsize]{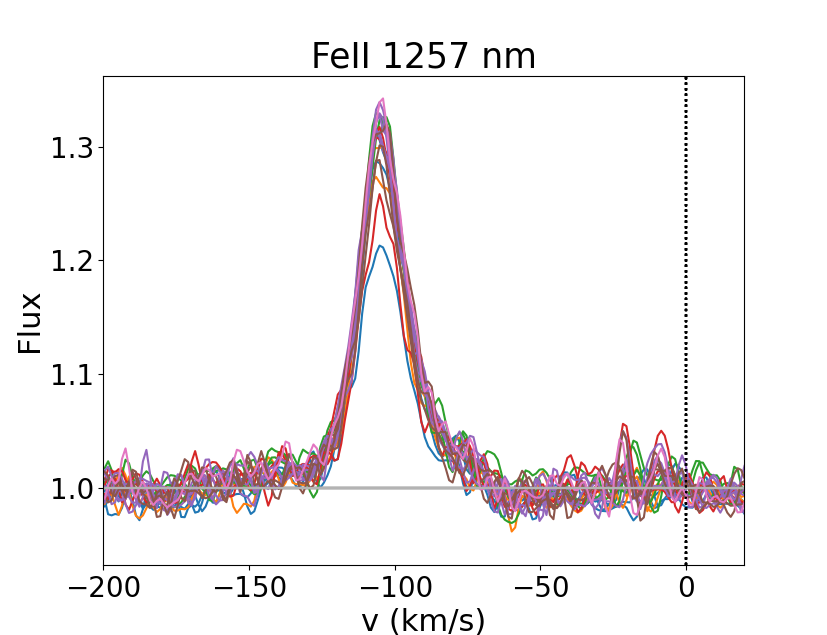}  
  \includegraphics[width=0.33\hsize]{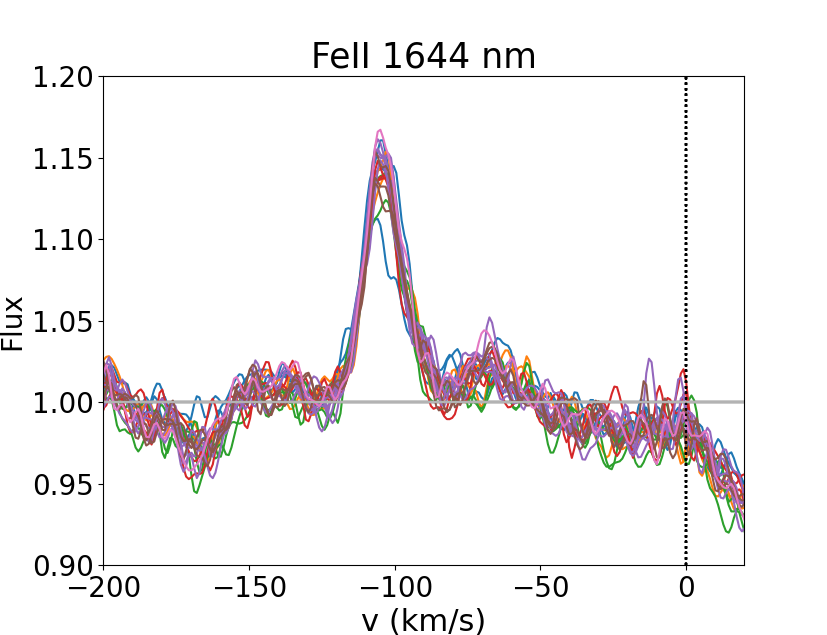}
  \includegraphics[width=0.33\hsize]{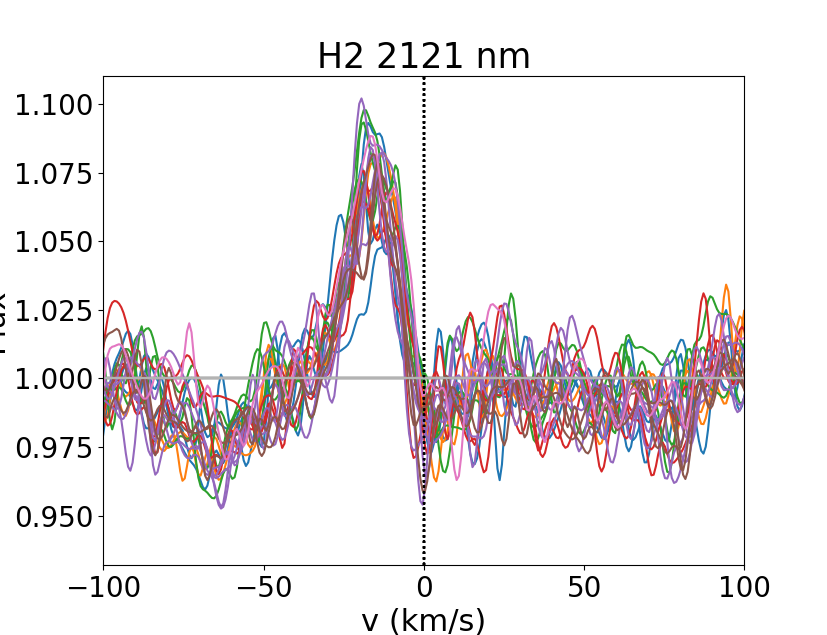} 
  \caption{Line profiles of Fe[II] 1257.02 and 1644.00 nm and H$_2$ 2121.83 nm from the SPIRou spectra. The color code corresponds to successive rotational cycles. The emission line profiles have not been corrected from underlying photospheric lines.}
  \label{fig:feiih2}%
\end{figure*}
\begin{figure*}[h]
    \centering
    \includegraphics[width=0.9\textwidth]{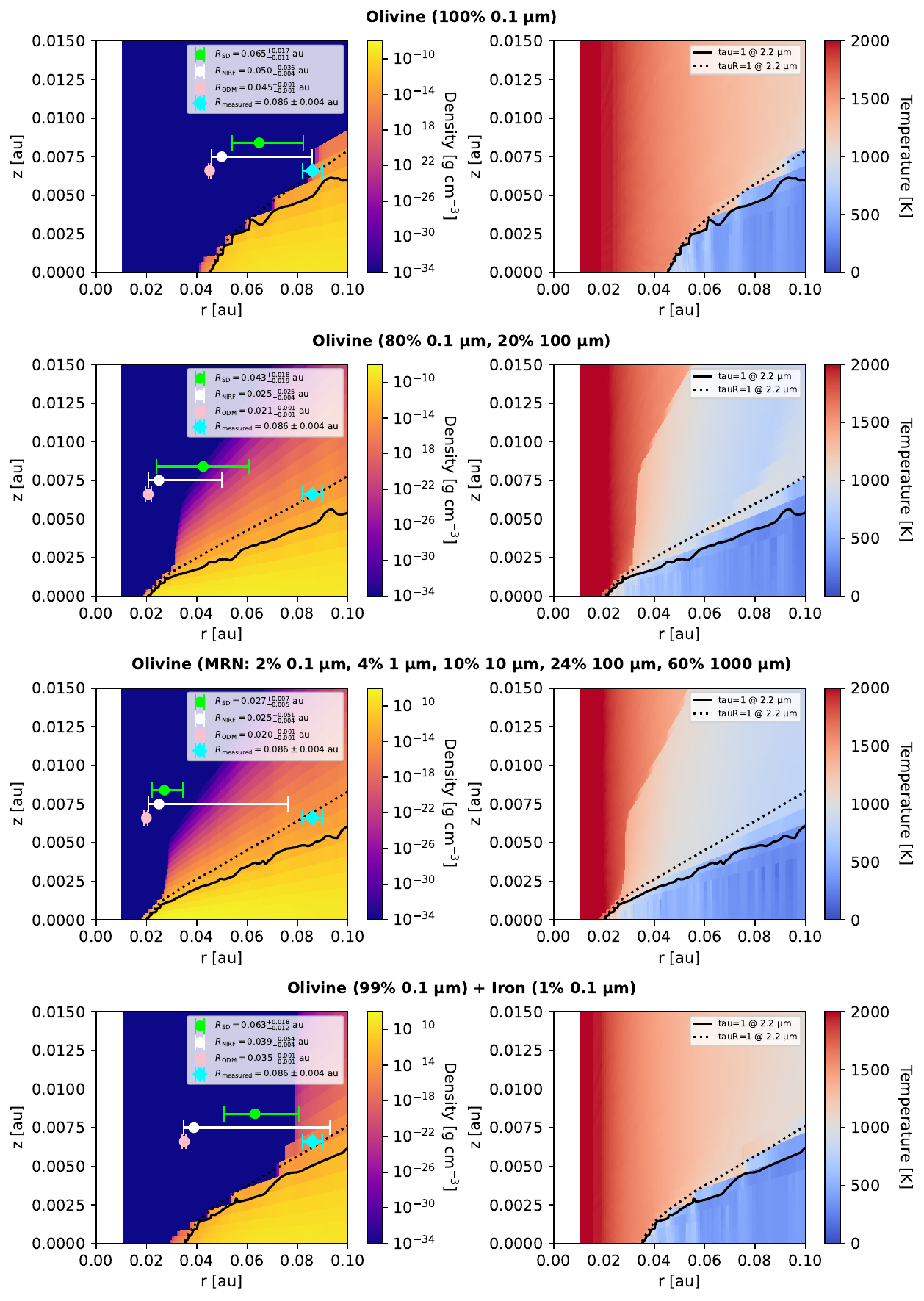}
    \caption{Density (left column) and temperature (right column) maps of the MCMax disk models for four different dust populations. The solid and dotted lines show the vertical and radial optically thin transitions at 2.2~µm, respectively. The green, white, and pink bars show the inner rim radii and widths extracted from the surface density, near-infrared flux, and optical depth at mid-plane, respectively. The cyan diamond and bar correspond to the GRAVITY measurement presented in this paper. The vertical position of the bars is arbitrary. }
    \label{fig:MCMaxmodels}
\end{figure*}

We used the amorphous olivine opacities of \cite{Dorschner1995} and iron opacities of \cite{Pollack1994}. Adopting the DIANA standard values of \cite{Woitke2016}, the grains have a porosity of 25\% and a maximum hollow volume of 80\%. Thermodynamic parameters for both materials come from \cite{Pollack1994}. After 150 iterations, following \cite{Klarmann2018} and taking into account the inclination of the star, we extracted the radii where the dust surface density reaches 10, 50, and 90\% of its peak ($R_{\rm SD}$, relevant for planetary formation); the radii inside which 10, 50, and 90\% of the 2.2~µm near-infrared flux is emitted ($R_{\rm NIRF}$, relevant for interferometric observations); and the radii where the optical depth in the midplane reaches 0.1, 1, and 10 ($R_{\rm ODM}$, relevant for radiative transfer modeling). All the values are gathered in Table~\ref{tab:radrim}.

The density and temperature maps from these simulations are shown in Fig.~\ref{fig:MCMaxmodels}, along with the modeled and measured radii and widths of the inner rim. When looking at $R_{\rm NIRF}$, the modeled radii range from 0.02 to 0.09~au, the closest radii with our measurement ($0.086\pm0.004$~au) being found for the small-grain models (pure olivine and olivine-iron mix). It is worth mentioning that the accretion luminosity is not taken into account in these models. This may push the inner rim radii further out and make the large-grain model more in line with the GRAVITY half-flux radius. 

\begin{table}[h]
\centering
\caption{Radius and width of the inner rim extracted from the surface density ($R_{\rm SD}$), near-infrared flux ($R_{\rm NIRF}$), and optical depth at midplane ($R_{\rm ODM}$) modeled with MCMax for four different dust populations. }
\label{tab:radrim}
\begin{tabular}{l|lll}
\hline \hline
Model                              & $R_{\rm SD}$ [au]          & $R_{\rm NIRF}$ [au]       & $R_{\rm ODM}$ [au]        \\
\hline
(1) Small grains           & $0.065^{+0.017}_{-0.011}$    & $0.050^{+0.036}_{-0.004}$   & $0.045^{+0.001}_{-0.001}$   \\ [1ex]
(2) Small/large grains          & $0.043^{+0.018}_{-0.019}$    & $0.025^{+0.025}_{-0.004}$   & $0.021^{+0.001}_{-0.001}$   \\ [1ex]
(3) MRN          & $0.027^{+0.007}_{-0.005}$    & $0.025^{+0.051}_{-0.004}$   & $0.020^{+0.001}_{-0.001}$   \\ [1ex]
(4) Olivine/iron grains          & $0.063^{+0.018}_{-0.012}$    & $0.039^{+0.054}_{-0.004}$   & $0.035^{+0.001}_{-0.001}$.  \\ [1ex]   
\hline
\end{tabular}
\end{table}

\end{appendix}

\end{nolinenumbers}
\end{document}